\documentclass[12pt,a4paper]{article}
\pdfoutput=1
\usepackage[utf8]{inputenc}
\newcommand{\sphid}[1]{}

\usepackage{epsf,amsmath,amssymb,graphicx,dcolumn}
\usepackage{caption}
\usepackage[labelformat=simple]{subcaption}

\usepackage{scalefnt,ulem,pstricks}
\usepackage{booktabs,multirow,tabularx}
\usepackage[colorlinks=true,allcolors={blue!70!black}]{hyperref}
\usepackage{cleveref}
\usepackage{color}
\usepackage{rotating}
\usepackage{microtype}
\usepackage[titletoc,title]{appendix}
\usepackage[numbers,sort&compress]{natbib}
\usepackage{amsmath,amsfonts,amsthm,bm}
\usepackage{xspace}
\providecommand{\href}[2]{#2}

\newcommand{\tmop}[1]{\ensuremath{\operatorname{#1}}}

\newcommand{\stepone}{{Step\,I}}
\newcommand{\steptwo}{{Step\,II}}
\newcommand{\stepthree}{{Step\,III}}

\newcommand\F{$Q\bar Q$}
\newcommand\FJ{$Q\bar Q{\rm J}$}
\newcommand\FJJ{$Q\bar Q{\rm JJ}$}

\newcommand{\as}{\alpha_s}

\newcommand{\pt}{{p_{\text{\scalefont{0.77}T}}}}

\newcommand{\ptrad}{{p_{\text{\scalefont{0.77}T,rad}}}}

\newcommand{\muF}{{\mu_{\text{\scalefont{0.77}F}}}}
\newcommand{\muR}{{\mu_{\text{\scalefont{0.77}R}}}}

\newcommand{\muRc}{{\mu_{\text{\scalefont{0.77}R}}^{(0)}}}
\newcommand{\KF}{{K_{\text{\scalefont{0.77}F}}}}
\newcommand{\KR}{{K_{\text{\scalefont{0.77}R}}}}

\newcommand{\noun}[1]{{\scshape #1}}

\newcommand{\POWHEG}{\noun{Powheg}}

\newcommand{\POWHEGBOXRES}{\noun{Powheg-Box-Res}}
\newcommand{\POWHEGBOXVTWO}{\noun{Powheg-Box-V2}}

\newcommand{\minlo}{{\noun{MiNLO$^{\prime}$}}\xspace}
\newcommand{\minnlo}{{\noun{MiNNLO$_{\rm PS}$}}\xspace}

\newcommand{\OpenLoops}{{\noun{OpenLoops}}\xspace}

\newcommand{\PYTHIA}[1]{\noun{Pythia{#1}}\xspace}

\newcommand{\setupatlas}{{\tt ATLAS}}
\newcommand{\setupcms}{{\tt CMS}}
\newcommand{\setuplhcbone}{{\tt LHCb-1}}
\newcommand{\setuplhcbtwo}{{\tt LHCb-2}}
\newcommand{\setuplhcbthree}{{\tt LHCb-3}}

\newcommand{\citere}[1]{Ref.\,\cite{#1}}

\newcommand{\citeres}[1]{Refs.\,\cite{#1}}

\newcommand{\eqn}[1]{Eq.\,(\ref{#1})}

\newcommand{\fig}[1]{Figure\,\ref{#1}}

\newcommand{\tab}[1]{Table\,\ref{#1}}

\newcommand{\LambdaPWG}{\Lambda_{\rm pwg}}

\usepackage{etoolbox}
\makeatletter
\patchcmd{\@sect}{#8}{\boldmath #8}{}{}
\let\ori@chapter\@chapter
\def\@chapter[#1]#2{\ori@chapter[\boldmath#1]{\boldmath#2}}
\makeatother

\usepackage{scalerel}

\usepackage[tikz]{bclogo}
\usepackage{subcaption}
\usepackage{tikz-feynman}
\usepackage{cancel}
\usepackage{verbatim}
\usetikzlibrary{arrows,shapes}
\usepackage{afterpage}

\begin{document} 
\begin{flushright}
\vspace*{-1.5cm}
MPP-2023-15 \\
PSI-PR-23-2
\end{flushright}
\vspace{0.cm}

\begin{center}
%{\Large \bf \boldmath{$B$}-hadron production at NNLO matched to parton showers}
%{\Large \bf Next-to-next-leading-order event generation \\[0.2cm] for \boldmath{$B$}-hadron production at the LHC}
%{\Large \bf NNLO+PS predictions for bottom-quark pair production}
%{\Large \bf Accurate description of \boldmath{$B$}-hadron production at the LHC \\[0.2cm] from NNLO+PS predictions for bottom-quark pairs}
{\Large \bf \boldmath{$B$}-hadron production at the LHC from\\[0.2cm] bottom-quark pair production at NNLO+PS}
\end{center}

\begin{center}
{\bf Javier Mazzitelli$^{(a)}$}, {\bf Alessandro Ratti$^{(b)}$}, \\[1ex]
{\bf Marius Wiesemann$^{(b)}$}, and {\bf Giulia Zanderighi$^{(b,c)}$}

(a) Paul Scherrer Institut, CH-5232 Villigen PSI, Switzerland

(b) Max-Planck-Institut f\"ur Physik, F\"ohringer Ring 6, 80805 M\"unchen, Germany

(c) Technische Universit\"at M\"unchen, James-Franck-Strasse 1, 85748 Garching, Germany

\href{mailto:javier.mazzitelli@psi.ch}{\tt javier.mazzitelli@psi.ch}\\
\href{mailto:ratti@mpp.mpg.de}{\tt ratti@mpp.mpg.de}\\
\href{mailto:marius.wiesemann@mpp.mpg.de}{\tt marius.wiesemann@mpp.mpg.de}\\
\href{mailto:zanderi@mpp.mpg.de}{\tt zanderi@mpp.mpg.de}

\end{center}

\begin{center} {\bf Abstract} \end{center}\vspace{-1cm}
\begin{quote}
  \pretolerance 10000
The production of $B$ hadrons is among the most abundant fundamental QCD processes 
measured at the LHC.
We present for the first time predictions for this process accurate to next-to-next-to-leading order
in QCD perturbation theory by simulating bottom-quark pair production at this accuracy 
matched to parton showers.
Our novel results are in good agreement with  experimental data  for the production of different types of $B$ hadrons 
from ATLAS, CMS and LHCb at 7\,TeV and/or 13 TeV, including various
fiducial cross sections as well as  single- and double-differential distributions, and 13\,TeV/7\,TeV cross-section 
ratios.

\end{quote}

\parskip = 1.2ex

\section{Introduction}

Precise simulations have become of foremost importance for the 
rich physics programme at the Large Hadron Collider (LHC), since
the experimental measurements are evolving at a substantial pace in terms
of statistical, and in some cases even systematical, 
uncertainties. Especially the lack of clear signals of new-physics 
phenomena calls for further precision studies at the LHC, which offer an
important pathway towards the discovery of physics beyond the Standard-Model (BSM) 
through small deviations from the Standard-Model (SM) predictions. In this context, the 
production of heavy quarks plays a fundamental role, as it provides a direct probe
of QCD interactions.

Bottom-quark pair ($b\bar{b}$) production is a particularly interesting and important process 
measured at the LHC in this class of processes. Although being the second heaviest
quark, bottom quarks are sufficiently light that they do not decay into 
elementary particles. Instead they directly form $B$ hadrons, which can be identified by 
the experiments as displaced vertices in their detectors, since $B$ hadrons
have a relatively long lifetime as their decay is strongly CKM suppressed.
The production of $B$ mesons or baryons originating from the hard $b\bar{b}$ process
has been extensively studied at hadron colliders: First measurements in 
proton--anti-proton collisions were already 
performed at CERN's Super Proton Synchrotron (SPS) by the UA1 collaboration
\cite{Albajar:1986iu,Albajar:1988th} and later at the Fermilab's Tevatron by CDF \cite{Abe:1995dv,Acosta:2001rz,Acosta:2004yw,Abulencia:2006ps} and D0 \cite{Abachi:1994kj,Abbott:1999se}. At the LHC,
all four experiments, including ALICE \cite{Abelev:2014hla,Abelev:2012sca}, ATLAS \cite{Aad:2012jga, ATLAS:2013cia}, CMS \cite{Khachatryan:2011mk,Chatrchyan:2011pw,Chatrchyan:2012hw,CMS:2016plw}, and LHCb \cite{Aaij:2010gn,Aaij:2012jd,LHCb:2013vjr,LHCb:2016qpe,LHCb:2017vec},
have presented  $B$-hadron measurements in proton--proton collisions at various 
centre-of-mass energies.

On the theoretical side, $B$-hadron production is an extremely interesting process.
Given that at typical LHC energies the bottom quarks, which are produced at the 
hard-process level, are right between being light or 
heavy quarks, there are essentially two relevant schemes to describe this process.
Either a massless description can be used in the so-called five-flavour scheme (5FS),
or the bottom quark can be treated as being massive in a four-flavour scheme (4FS).
On the other hand, for a realistic description of $B$ mesons and baryons measured by the 
experiments it is necessary to include hadronization effects on top of an accurate 
description of bottom-quark kinematics. This can be achieved by combining a fixed-order
calculation of $b\bar{b}$ production with parton showers that include hadronization
models. At least, a consistent inclusion of fragmentation functions to account for relevant 
resummation effects is necessary to cover all  phase-space regions.

Higher-order corrections to this process are particularly important as the  
small bottom mass leads to a relatively small natural scale that in turn implies
large values of the strong coupling and a slow convergence of the QCD series.
Next-to-leading order (NLO) corrections in QCD are 
known since a long time \cite{Nason:1987xz,Nason:1989zy,Beenakker:1988bq,Mangano:1991jk},
while more recently uncertainties related to the 
renormalization of the bottom-quark mass have been addressed \cite{Garzelli:2020fmd}.
Resummation effects, relevant already at rather moderately large transverse momenta, 
have been included in different 
approaches \cite{Cacciari:1993mq,Mele:1990cw,Cacciari:2001cw,Kniehl:2004fy, Kniehl:2005mk,Kramer:2018vde,Benzke:2019usl}.
Their combination with NLO QCD corrections, dubbed ``FONLL'' \cite{Cacciari:1998it, Cacciari:2001td,
Cacciari:2002pa, Cacciari:2012ny}, in combination with non-perturbative fragmentation functions\footnote{So far, fragmentation functions are typically extracted from LEP data \cite{Cacciari:2005uk}.}
\cite{Kartvelishvili:1977pi, Peterson:1982ak}
has been the reference prediction in experimental analyses for a long time.
The combination of NLO QCD predictions with parton showers has been achieved 
  with various schemes and implemented in various tools~\cite{Frixione:2007nw,Buonocore:2017lry,Alwall:2014hca}.
These calculations, enable a fully realistic description of $B$ hadrons 
at the level of fully exclusive events in hadronic collisions while keeping NLO QCD accuracy.
On the other hand, it has been shown that also the next-to-NLO (NNLO) QCD 
predictions are indispensable for $b\bar{b}$ production, as they lead to substantial 
corrections both for the total 
inclusive rate \cite{Mangano:2016jyj,dEnterria:2016ids}, as 
 implemented in the numerical code {\sc Hathor} \cite{Langenfeld:2009wd,Aliev:2010zk},
and fully differentially in the kinematics of the bottom quarks \cite{Catani:2020kkl}, which is implemented in the {\sc Matrix} framework \cite{Grazzini:2017mhc}.\footnote{Recently, there
has been a NNLO QCD description of $B$ hadrons originating from top-quark decays in top-quark pair
production~\cite{Czakon:2021ohs}.}

In this letter we achieve 
the first simulation of $B$-hadron production in hardonic collisions
at NNLO in QCD. To this end, we calculate 
NNLO QCD corrections to $b\bar{b}$ 
production and, for the first 
time, consistently combine them with parton showers.
This allows us to obtain hadron-level events keeping NNLO QCD accuracy 
and including the hadronization of the bottom quarks.
Our calculation follows closely the corresponding simulation for 
top-quark pair production \cite{Mazzitelli:2020jio,Mazzitelli:2021mmm}. 
We perform a validation against 
fixed-order NNLO QCD results and present an extensive comparison 
against $7$ and $13$ TeV results of different LHC experiments,
where we find that our predictions are in remarkably good agreement with 
the measurements.

\section{Outline of the calculation}

\begin{figure}[t]
  \begin{center}
    \begin{subfigure}[b]{.3\linewidth}
      \centering
\begin{tikzpicture}
\begin{feynman}
	\vertex (a1) at (0,0) {\( q\)};
	\vertex (a2) at (0,-2) {\(\bar q\)};
	\vertex (a3) at (1.4,-1);
	\vertex (a4) at (2.7,-1);
	\vertex (a5) at (4.1,0){\( b\)};
	\vertex (a6) at (4.1,-2){\(\bar b\)};
        \diagram* {
          {[edges=fermion]
            (a1)--[thick](a3)--[thick](a2),
            (a6)--[fermion, ultra thick](a4)--[fermion, ultra thick](a5),
          },
          (a3) -- [gluon,thick] (a4),
        };
      \end{feynman}
\end{tikzpicture}
\caption{$s$-channel $q\bar{q}$ diagram}
        \label{subfig:qq}
\end{subfigure}%
\begin{subfigure}[b]{.3\linewidth}
  \centering
\begin{tikzpicture}
  \begin{feynman}
	\vertex (a1) at (0,0) {\( g\)};
	\vertex (a2) at (0,-2) {\( g\)};
	\vertex (a3) at (1.4,-1);
	\vertex (a4) at (2.7,-1);
	\vertex (a5) at (4.1,0){\( b\)};
	\vertex (a6) at (4.1,-2){\(\bar b\)};
        \diagram* {
          {[edges=fermion]
            (a6)--[fermion, ultra thick](a4)--[fermion, ultra thick](a5),
          },
          (a3) -- [gluon,thick] (a4),
          (a2)--[gluon,thick](a3)--[gluon,thick](a1),
        };
  \end{feynman}
\end{tikzpicture}
\caption{$s$-channel $gg$ diagram}
        \label{subfig:gg}
\end{subfigure}%
\begin{subfigure}[b]{.3\linewidth}
  \centering
\begin{tikzpicture}
  \begin{feynman}
	\vertex (a1) at (0,0) {\( g\)};
	\vertex (a2) at (0,-1.7) {\( g\)};
	\vertex (a3) at (1.53,0);
	\vertex (a4) at (1.53,-1.7);
	\vertex (a5) at (3,0){\( b\)};
	\vertex (a6) at (3,-1.7){\(\bar b\)};
        \diagram* {
          {[edges=fermion]
            (a6)--[fermion, ultra thick](a4)--[fermion, ultra thick](a3)--[fermion, ultra thick](a5),
          },
          (a2)--[gluon,thick](a4),
          (a3)--[gluon,thick](a1),
        };
  \end{feynman}
\end{tikzpicture}\vspace{0.15cm}
\caption{$t$-channel $gg$ diagram}
        \label{subfig:gg}
\end{subfigure}
\end{center}
\caption{\label{fig:diagrams} Feynman diagrams for 
  the process $pp\to b\bar{b}$ at LO.}
\end{figure}
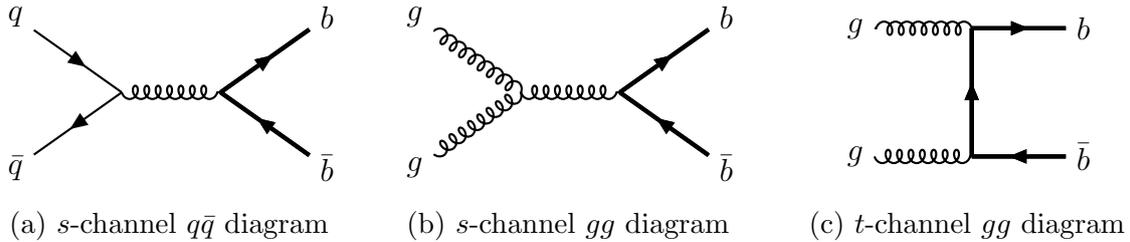

We are interested in obtaining theoretical predictions for the process
\begin{align}
\label{eq:proc}
pp\to B + X\,,
\end{align}
where $B$ is a hadron containing either a bottom or an anti-bottom quark, but not both, and $X$ indicates that we are otherwise inclusive over the final state. Note that the experimental analysis typically focus on mesons  ($B^0,\bar B^0,B^+,B^-,B_s^0,\bar B_s^0,\ldots$), 
or a subset of them as we shall see later, while baryons ($\Lambda_b^0,\bar\Lambda_b^0,\Xi_b^0,\Xi_b^-,\Omega_b^-,\dots$) are only included in some analyses. 
Moreover, the $B$ hadrons listed above are the ones included in the most inclusive $B$-hadron measurements, since, for instance, the production
of $B_c$ mesons yields only a negligible fraction ($\sim 0.1\%$) of all $B$ hadrons.

To simulate the process in \eqn{eq:proc} we have implemented a fully differential computation of $pp\to b\bar{b}$ production
to NNLO in the expansion of the strong coupling constant 
and consistently matched it to a parton-shower simulation (NNLO+PS), which allows us to generate exclusive events with hadronic final states.
Leading-order (LO) Feynman diagrams for this process are shown in \fig{fig:diagrams}. 
Our calculation is based on the \minnlo{} method \cite{Monni:2019whf,Monni:2020nks}, 
which was originally developed and later used for several colour-singlet processes \cite{Lombardi:2020wju,Lombardi:2021rvg,Buonocore:2021fnj,Lombardi:2021wug,Zanoli:2021iyp,Gavardi:2022ixt,Haisch:2022nwz,Lindert:2022qdd}. 
More precisely, we employ the extension of the \minnlo{} method that was derived and applied to top-quark pair
production in \citeres{Mazzitelli:2020jio,Mazzitelli:2021mmm} in order to implement a NNLO+PS generator for bottom-quark pair production. 

We briefly recall the basics of the \minnlo{} approach for heavy-quark pair production and we
refer the interested reader to \citere{Mazzitelli:2021mmm} for the complete description of the method
and its detailed derivation. 
For the sake of brevity, we adopt a rather simplified and symbolic notation here.
The \minnlo{} formalism allows us to include NNLO corrections in the event generation
of a heavy quark pair ($Q\bar Q$). It is derived from the analytical transverse-momentum 
resummation formula for $Q\bar Q$ production \cite{Zhu:2012ts,Li:2013mia,Catani:2014qha,Catani:2018mei}, which captures the logarithmic terms up to a given perturbative 
order and with a certain logarithmic accuracy. After appropriate simplifications, which are allowed 
within our desired accuracy (i.e.\ NNLO and preserving the accuracy of the parton shower), the 
relevant singular terms in the transverse momentum of the $Q\bar Q$ pair
can be written in the following symbolic form \cite{Mazzitelli:2020jio,Mazzitelli:2021mmm}:
\begin{align}
\label{eq:resum}
{\rm d}\sigma_{\scriptscriptstyle Q\bar Q}^{\rm res}&=\frac{{\rm d}}{{\rm d}\pt}\left\{\left[\sum_{i=1}^{n_c}\mathcal{C}_i\,e^{-S_i}\right]\mathcal{L}\right\}=\sum_{i=1}^{n_c}\mathcal{C}_i\,e^{-S_i}\underbrace{\left\{-S_i^\prime\,\mathcal{L}+\mathcal{L}^\prime\right\}}_{\equiv D_i}\,.
\end{align}
Note that the sum over partonic channels shall be understood as being implicit.
Moreover, in contrast to the colour-singlet case, an explicit sum over $n_c$ appears ($n_c=4$ for $q\bar{q}$ channels and $n_c=9$ for the $gg$ channel), 
which originates from independent colour configurations. This notation is required, since the 
logarithmic corrections arising from soft wide-angle exchanges between the final-state heavy quarks as well as final-initial state interferences
render the soft anomalous dimensions for heavy-quark pair production $\mathbf{\Gamma}^{(1)}_t$ to be matrix/operator in colour space. Thus,
through its exponentiation the Sudakov form factor becomes colour dependent as indicated by the subscript $i$.
The sum in $i$ is a consequence of diagonalizing $\mathbf{\Gamma}^{(1)}_t$, which generates the complex coefficients $\mathcal{C}_i$ 
that fulfil  $\sum_{i=1}^{n_c}\mathcal{C}_i=1$. The eigenvalues of  $\mathbf{\Gamma}^{(1)}_t$ are included in a redefinition
of the $B^{(1)}$ coefficient of the Sudakov form factor.
In addition to that, also the $B^{(2)}$ coefficient
 is modified such as to reproduce all singular terms up to NNLO correctly,
by including the contributions from 
$\mathbf{\Gamma}^{(2)}_t$ and by compensating for the approximation 
that $\mathbf{\Gamma}^{(1)}_t$ is diagonalized with the LO colour-decomposed hard-scattering amplitude.
We refer to \citere{Mazzitelli:2021mmm} for details, in particular to Eq.\,(3.20) of  that paper.
We also stress that the luminosity factor $\mathcal{L}$, which 
includes the squared hard-virtual matrix elements for $Q\bar Q$ production and the convolution of the 
collinear coefficient functions with the parton distribution functions (PDFs),
includes additional contributions from soft wide-angle exchanges between the heavy quarks and with 
the initial state as well, obtained from the calculation presented in \citere{Catani:2023tby}.
In particular, these induce azimuthal correlations and require us to include
additional contributions after taking the azimuthal average, see Eq.\,(3.31) of \citere{Mazzitelli:2021mmm}.

Apart from these (subtle, but crucial) modifications of the singular contributions, their general structure in \eqn{eq:resum} is 
very reminiscent of the colour-singlet case. Thus, while keeping the information on the different colour configurations explicit, 
we can now follow the same procedure to derive a formula to construct a NNLO+PS generator for $Q\bar Q$ production.
To this end, we combine the singular terms (up to NNLO in QCD) in \eqn{eq:resum} with the 
the differential cross section of a heavy-quark pair and a jet (\FJ{}) at  
first- and second-order, ${\rm d}\sigma^{(1,2)}_{\scriptscriptstyle Q\bar Q{\rm J}}$, while removing 
any double counting and using a
matching scheme where the Sudakov form factor is factored out:
\begin{align}
 & {\rm d}\sigma_{\scriptscriptstyle Q\bar Q}^{\rm res}+[{\rm d}\sigma_{\scriptscriptstyle Q\bar Q{\rm J}}]_{\rm f.o.}-[{\rm d}\sigma_{\scriptscriptstyle Q\bar Q}^{\rm res}]_{\rm f.o.}=\sum_{i=1}^{n_c}\mathcal{C}_i\,e^{-S_i}\bigg\{D_i+[{\rm d}\sigma_{\scriptscriptstyle Q\bar Q{\rm J}}]_{\rm f.o.}\,\underbrace{\frac{1}{[e^{-S_i}]_{\rm f.o.}\,}}_{1+S_i^{(1)}\cdots}\underbrace{\,\,-\,\frac{[{\rm d}\sigma_{\scriptscriptstyle Q\bar Q}^{\rm res}]_{\rm f.o.}\,\,}{[e^{-S_i}]_{\rm f.o.}}}_{-D_i^{(1)}-D_i^{(2)}\cdots}\bigg\}\\
&\quad\quad\quad\quad\quad\quad\quad\,\,\approx \sum_{i=1}^{n_c}\mathcal{C}_i\,e^{-S_i}\,\bigg\{{\rm d}\sigma^{(1)}_{\scriptscriptstyle Q\bar Q{\rm J}}\big(1+S_i^{(1)}\big)+{\rm d}\sigma^{(2)}_{\scriptscriptstyle Q\bar Q{\rm J}}+\underbrace{\left(D_i-D_i^{(1)}-D_i^{(2)}\right)}_{\equiv D_i^{(\ge 3)}}\bigg\}\,,\nonumber
\end{align}
where $[\cdots]_{\rm f.o.}$ denotes the expansion up to a given fixed order in $\as$,
and $X^{(n)}$ is the $n$-th coefficient in the $\as$ expansions of $X$ including 
the coupling $\as^n$ itself. Note that, in the last step, we have neglected terms beyond NNLO QCD accuracy for inclusive \F{} production.
We can apply this exact 
procedure directly in a \POWHEG{} \cite{Nason:2004rx,Nason:2006hfa,Frixione:2007vw,Alioli:2010xd} calculation 
for the \FJ{} process to obtain the \minnlo{} master formula for heavy-quark pair production:
\begin{align}
\label{eq:master}
{\rm d}\sigma_{\scriptscriptstyle Q\bar Q}^{\rm MiNNLO_{PS}}={\rm d}\Phi_{\scriptscriptstyle Q\bar Q{\rm J}}\,\bar{B}^{\,\rm MiNNLO_{\rm PS}}\,\times\,\left\{\Delta_{\rm pwg}(\Lambda_{\rm pwg})+ {\rm d}\Phi_{\rm rad}\Delta_{\rm pwg}(p_{T,{\rm rad}})\,\frac{R_{\scriptscriptstyle Q\bar Q{\rm J}}}{B_{\scriptscriptstyle Q\bar Q{\rm J}}}\right\}\,,
\end{align}
where the standard \POWHEG{} $\bar B$ function is modified as 
\begin{align}
\label{eq:minnlo}
\bar{B}^{\,\rm MiNNLO_{\rm PS}}\sim \sum_{i=1}^{n_c}\mathcal{C}_i\,e^{-S_i}\,\bigg\{{\rm d}\sigma^{(1)}_{\scriptscriptstyle Q\bar Q{\rm J}}\big(1+S_i^{(1)}\big)+{\rm d}\sigma^{(2)}_{\scriptscriptstyle Q\bar Q{\rm J}}+D_i^{(\ge 3)}\times F^{\rm corr}\bigg\}\,,
\end{align}
ensuring NNLO QCD accuracy for \F{} production when the additional jet becomes unresolved.
With $\Phi_{\scriptscriptstyle Q\bar Q{\rm J}}$ we denote the \FJ{} phase space, with $\Delta_{\rm pwg}$ the \POWHEG{} Sudakov form factor
featuring a default cutoff of $\LambdaPWG=0.89$\,GeV, and with $\Phi_{\tmop{rad}} $ and $\ptrad$ 
the phase space and the transverse momentum of the second radiation. $B_{\scriptscriptstyle Q\bar Q{\rm J}}$ and $R_{\scriptscriptstyle Q\bar Q{\rm J}}$ are the squared tree-level matrix elements for \FJ{} and \FJJ{} production, respectively.
NNLO QCD accuracy is achieved through the third term in \eqn{eq:minnlo}, which adds the relevant (singular) contributions of order $\as^3(p_{\text{\scalefont{0.77}T}})$~\cite{Monni:2019whf}. Regular contributions at this order are of subleading nature. Moreover, \minlo{} results, which
correspond to a merging of $0$-jet and $1$-jet multiplicities at NLO QCD accuracy, are defined  
by not including the NNLO $D_i^{(\ge 3)}$ corrections in \eqn{eq:minnlo}.\footnote{Note that the \minlo{} predictions for $b\bar{b}$ production are also an original result of the present paper.}

A few comments are in order: In \eqn{eq:minnlo}, the extra real radiation with respect
to the \FJ{} process (i.e. \FJJ{}), including its phase space and standard \POWHEG{}
mappings, is implicit,
 and similarly an appropriate projection from the \FJ{} to the \F{} phase-space is understood,
where the factor $F^{\rm corr}$ encodes the appropriate function which ensures that the NNLO corrections are spread 
in the \FJ{} phase space. This spreading function is a necessary ingredient for the implementation of the NNLO corrections to \F{} production in the context of a \FJ{} \POWHEG{} implementation.
With a slight abuse of notation, the transverse momentum of the $Q\bar Q$ pair $p_T$ for the evaluation of the Sudakov form factor 
is determined in each respective phase space \FJ{} or \FJJ{}, respectively. As stated before, the sum over all flavour configurations,
especially with respect to \FJ{} and \FJJ{} configurations (and where appropriate to \F{}),
is understood implicitly in \eqn{eq:minnlo} as well, and so is the appropriate projection of the flavour configuration to 
decide whether the $q\bar{q}$ or $gg$ Sudakov is used in a specific \FJ{} or \FJJ{}  flavour configuration. In particular in the  $qg$/$gq$-initiated  channels 
 we follow the procedure described in \citere{Mazzitelli:2021mmm}.

The essential steps behind the \minnlo{} procedure can be summarized as follows: 
in the first one (\stepone{}) the \FJ{} final state is described
at NLO QCD accuracy using \POWHEG{}, inclusively over the radiation of a second light parton.
The second step (\steptwo{}), which characterizes the \minnlo{} approach,
appropriately regulates the limit in which the light partons become unresolved 
by supplementing the correct Sudakov behaviour as well as higher-order terms, 
such that the simulation becomes NNLO QCD accurate for inclusive
\F{} production.
These first two steps are included in the $\bar{B}$ function of \eqn{eq:minnlo}.
In the third step (\stepthree{}), the second radiated parton is generated exclusively (accounted for inclusively in \stepone{}) 
through the content of the curly brackets in \eqn{eq:master}, keeping the
NLO (and NNLO) QCD accuracy of \FJ{} (and \F{}) production untouched, while
subsequent radiation is included through the parton shower.

In these three steps all emissions are appropriately ordered
(when using a $p_T$-ordered shower) and the applied Sudakov matches
the logarithmic structure of the parton shower.  As a result, the
\minnlo{} approach preserves the (leading logarithmic) 
accuracy of the parton shower, while reaching NNLO QCD accuracy in 
the event generation.
Besides these physically essential features, we recall that what
  makes \minnlo{} such a powerful approach is that the event
  simulation is extremely efficient, which is a result of two facts.
First, NNLO QCD corrections are calculated directly during the 
event generation (no need for any a-posteriori reweighting). Second, 
due to the appropriate suppression through the Sudakov form factor 
no unphysical merging scale or slicing cutoff is required to separate 
different multiplicities in the generated event samples. Not only does this
ensure that there are no large cancellation between different contributions,
it also keeps all power-suppressed terms into account, as opposed to 
approaches that resort to non-local/slicing techniques.
Finally, we stress that, although the \minnlo{} method has been initially 
developed on the basis of the transverse momentum of the colour singlet and later
been rederived for the transverse momentum of a heavy-quark pair, it is 
clear that the main idea behind the approach is neither limited to a specific observable,  
nor to these two classes of processes.

In contrast to the computation for top-quark pair ($t\bar t$) production in \citeres{Mazzitelli:2020jio,Mazzitelli:2021mmm}, 
our calculation is performed in the \POWHEGBOXRES{} framework \cite{Jezo:2015aia}
instead of \POWHEGBOXVTWO{} \cite{Alioli:2010xd}. To this end, we have exploited the interface to \OpenLoops{}~\cite{Cascioli:2011va,Buccioni:2017yxi,Buccioni:2019sur}, 
which was developed in \citere{Jezo:2016ujg}, to obtain the $pp\to b\bar{b}$+jet process at NLO+PS in the 4FS with massive bottom quarks (used throughout in our calculation).
To reach NNLO+PS accuracy for $pp\to b\bar{b}$ production we have performed a new implementation 
of the NNLO+PS method outlined above for heavy-quark pair production within \POWHEGBOXRES{} based on \minnlo{}.
The tree-level and one-loop amplitudes are therefore evaluated through  \OpenLoops{}, while for the two-loop amplitude 
we rely on the numerical implementation of \citere{Barnreuther:2013qvf}.
As an important cross check, we have implemented not only a new generator for bottom-quark pair
production, but also for top-quark pair production in \POWHEGBOXRES{} and verified that we 
find full agreement within numerical uncertainties when compared to our original $t\bar t$ code in 
\POWHEGBOXVTWO{} \cite{Mazzitelli:2020jio,Mazzitelli:2021mmm}.

\section{Phenomenological Results}

We now turn to presenting phenomenological results for $b\bar{b}$ production and $B$-hadron production 
at the LHC at different centre-of-mass energies. Besides a fully inclusive setup that is used for validation purposes, 
we compare our NNLO+PS predictions against four different experimental measurements: a 7\,TeV measurement by ATLAS \cite{ATLAS:2013cia} and LHCb  \cite{LHCb:2013vjr}
referred to as \setupatlas{} and \setuplhcbone{}, respectively; two different 
LHCb analyses that contain both 7\,TeV and 13\,TeV measurements as well as their ratios denoted in the following 
as \setuplhcbtwo{} \cite{LHCb:2016qpe} and \setuplhcbthree{} \cite{LHCb:2017vec}, respectively; and a 13\,TeV measurement by CMS \cite{CMS:2016plw} referred to as \setupcms{}.
We refer to those publications for the definition of the respective fiducial phase spaces and we note that 
all of these measurements involve a different selection of $B$ hadrons in their analyses, which will be detailed below.

Our calculation employs the 4FS throughout. Therefore, the bottom quarks are treated as being massive and we set their 
pole mass to $m_b=4.92$\,GeV. For the PDFs we choose the NNLO set of the 
NNPDF3.1~\cite{Ball:2017nwa} consistent with $N_f=4$ number of light quark flavours (specifically {\tt NNPDF31\char`_nnlo\char`_as\char`_0118\char`_nf\char`_4})
and the strong coupling with 4FS running corresponding to that set.
The PDFs are read via the {\sc lhapdf} interface \cite{Buckley:2014ana}, but they are copied to the {\sc hoppet} code \cite{Salam:2008qg} that performs their internal evolution and the relevant convolutions.
The setting of the renormalization~($\muR$) and factorization ($\muF$) scales for our default \minnlo{} 
and \minlo{} predictions
 is fixed by the method itself and described in section 4.3 of \citere{Mazzitelli:2021mmm}, with the only exception of the scale entering the two powers of $\as$ that are already present at Born level.
We also follow the definition of the modified logarithm of that paper to consistently switch off 
resummation effects at large transverse momenta with the standard scale choice of $Q=m_{b\bar b}/2$.
All other technical settings are kept as in \citere{Mazzitelli:2021mmm} as well ($Q_0 = 2$\,GeV, $\KR = \KF = 1$ for the central scales). 
The scales for the two overall powers of $\as$ at Born level are set to
\begin{align}
\muRc=\KR\,\frac{H_T^{b\bar b}}2,\quad\text{with}\quad H_T^{b\bar b}= \sqrt{m_b^2+p_{T,b}^2}+\sqrt{m_b^2+p_{T,\bar b}^2}\,.
\end{align}
For validation purposes we compare against fixed-order NNLO predictions computed at the scale $\muR=\KR\,m_{b\bar b}$, $\muF=\KF\,m_{b\bar b}$. Therefore, only in this case we use $\muRc=\KR\,m_{b\bar b}$ in our \minnlo{} implementation instead, to provide a more direct comparison. 
In all cases we use $7$-point scale variations, i.e.\ varying $\KR$ and $\KF$ by a factor of two in each direction
with the constraint $1/2\le \KR/\KF\le 2$, 
to estimate the uncertainties related to missing higher-order contributions.
All showered results have been obtained with \PYTHIA{8} \cite{Sjostrand:2014zea} with the Monash 2013 tune \cite{Skands:2014pea},
where we have turned on effects from  hadronization and from multi-parton interactions (MPI)  to obtain a fully realistic simulation of $B$ hadrons.

\begin{figure}[t!]
\begin{center}
\begin{tabular}{ccccc}
%\includegraphics[width=.34\textwidth, page=1]{plots/validation_plots.pdf}
%&
%\hspace{-0.55cm}
%\includegraphics[width=.34\textwidth, page=3]{plots/validation_plots.pdf}
%&
%\hspace{-0.55cm}
%\includegraphics[width=.34\textwidth, page=5]{plots/validation_plots.pdf}\\
\hspace{-0.5cm}
\includegraphics[width=.34\textwidth, page=3]{plots/validation_average.pdf}
&
\hspace{-0.55cm}
\includegraphics[width=.34\textwidth, page=1]{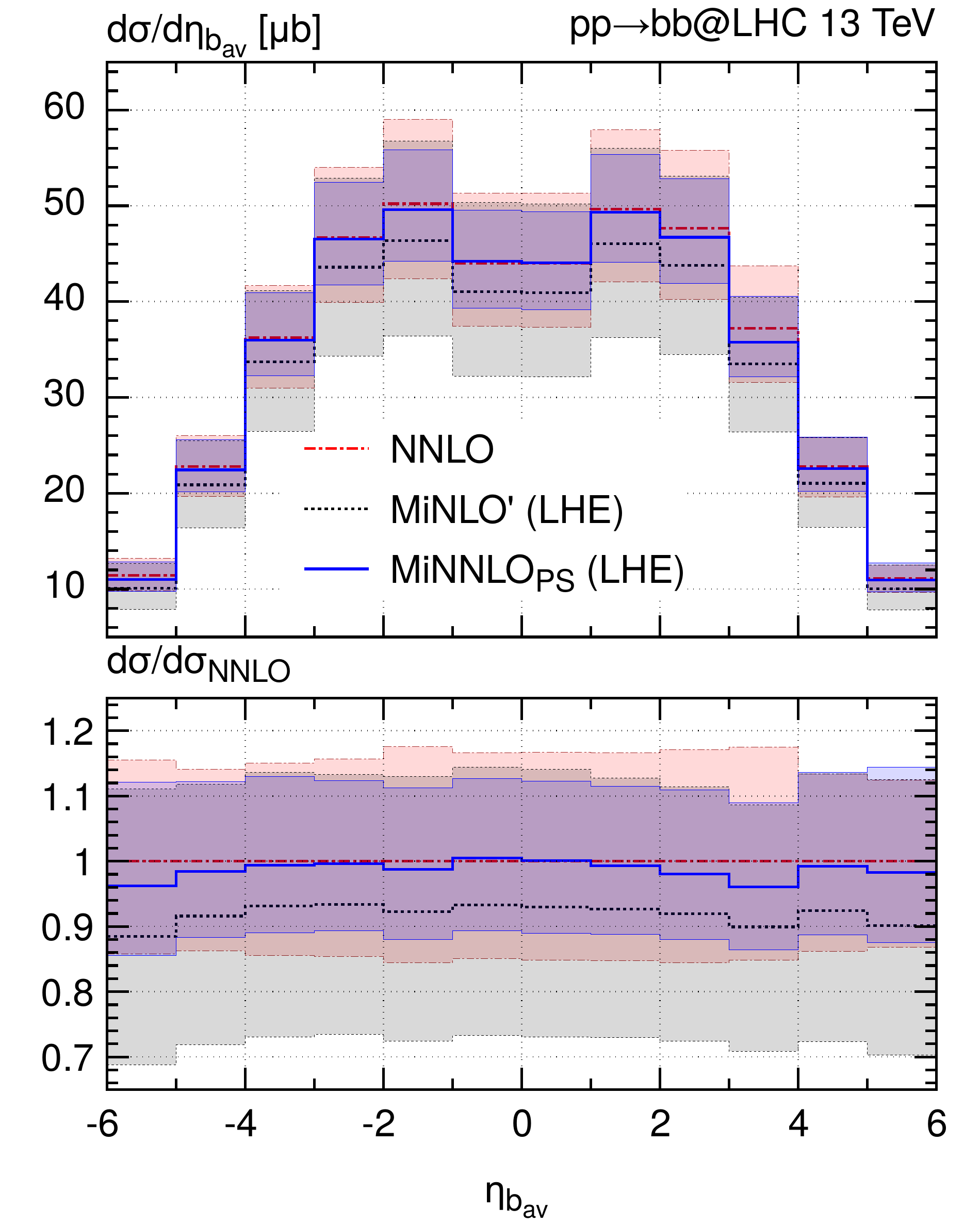}
&
\hspace{-0.55cm}
\includegraphics[width=.34\textwidth, page=2]{plots/validation_average.pdf}
\end{tabular}
\vspace*{1ex}
\caption{\label{fig:validation} Validation of \minnlo{} predictions against NNLO QCD results. See text for details.}
\end{center}
\end{figure}

We start by validating the predictions of our \minnlo{} generator in the fully inclusive phase space of the bottom-quark pair 
against fixed-order NNLO results \cite{Catani:2020kkl} from {\sc Matrix} \cite{Grazzini:2017mhc}.
The fully inclusive \minnlo{} cross section for bottom-quark pair production
amounts to $428.7(6)_{-11\%}^{+13\%}\,\mu b$, which is in perfect agreement with the fixed-order 
NNLO QCD result of $435(2)_{-15\%}^{+16\%}\,\mu b$.
For a more direct comparison, the \minnlo{} distributions are shown 
here at the Les-Houches-Event (LHE) level without showering effects.
\fig{fig:validation} displays three differential distributions in the kinematics of the
bottom quarks 
 namely the average rapidity ($y_{b_{\rm av}}$), pseudo-rapidity ($\eta_{b_{\rm av}}$), and transverse momentum ($p_{T,b_{\rm av}}$) of the bottom and antibottom quark.
These observables are all non-trivial in the Born phase space and therefore they are genuinely NNLO QCD accurate. Besides \minnlo{} (blue solid) and
fixed-order NNLO QCD predictions (red dashed), we also include \minlo{} results (black dotted) as a reference, which are NLO QCD accurate 
in all 0-jet (and 1-jet) observables. First of all, we see in \fig{fig:validation} that the NNLO QCD corrections included through the \minnlo{} procedure have an impact of 
$\mathcal{O}(+10\%)$ with respect to \minlo{}. Moreover, the corrections are very flat as a function of the rapidities, while there are slight shape effects in $p_{T,b_{\rm av}}$.
However, one should bear in mind that the 0-jet and 1-jet merged \minlo{} prediction already includes
important corrections (beyond fixed-order NLO QCD) due to the NLO QCD corrections to hard parton radiation
in the 1-jet phase space. When comparing \minnlo{} predictions with the fixed-order NNLO QCD results, we find
that all observables are fully compatible within their respective perturbative uncertainties. We note that
these two calculations differ by terms beyond accuracy and they are not expected to yield identical results. 
Despite this fact, one can see that the central predictions are extremely close and the size of the uncertainty bands
is very similar. With this we conclude the validation of the NNLO QCD accuracy of the \minnlo{} generator
and move on to considering predictions for $B$ hadron production at the LHC.

\renewcommand*{\arraystretch}{1.3}
\begin{table}[t]
\begin{center}
\resizebox{\textwidth}{!}{  
    \begin{tabular}{ |c|c|c|c|c| }
    \hline

    Analysis
    &
    Energy
    &
    Process
    &
    Measured cross section ($\mu b$)
    &
    %$\sigma^{\rm MiNNLO_{\rm PS}}$ 
    \minnlo{}  ($\mu b$)
\\
    \hline
    \hline
    \multirow{1}{*}{\setupatlas{} \cite{ATLAS:2013cia}}
    &
    \multirow{1}{*}{7\,TeV}
    &
    \multirow{1}{*}{$pp \rightarrow B^{+}$+$X$}
    & $10.6 \pm 0.3\text{\scriptsize (stat)} \pm 0.7\text{\scriptsize (syst)}$ $\pm 0.2\text{\scriptsize (lumi)} \pm 0.4\text{\scriptsize (bf)}$
    & \multirow{1}{*}{$10.17(5)_{-14.0\%}^{+13.3\%}$}
        \\
    \hline
    \setupcms{} \cite{CMS:2016plw} &
    13\,TeV
     & $pp \rightarrow B^{+}$+$X$
    & $15.3 \pm 0.4\text{\scriptsize (stat)} \pm 2.1\text{\scriptsize (syst)} \pm 0.4\text{\scriptsize (lumi)}$
    & $11.47(6)_{-13.2\%}^{+11.3\%}$
        \\
    \hline
    \multirow{3}{*}{\setuplhcbone{} \cite{LHCb:2013vjr}} &
    \multirow{3}{*}{7\,TeV}
    &
    $pp \rightarrow B^{\pm}$+$X$
    & $38.9\pm 0.3\text{\scriptsize (stat)} \pm 2.5\text{\scriptsize (syst)} \pm 1.3\text{\scriptsize (bf)}$
    & ${42.2(1)}_{-11.4\%}^{+13.9\%}$
    \\ 
    &
    &
    $pp \rightarrow B^{0}$+$X$
    & $38.1 \pm 0.6\text{\scriptsize (stat)} \pm 3.7\text{\scriptsize (syst)} \pm 4.7\text{\scriptsize (bf)}$
    & $42.3(1)_{-11.3\%}^{+14.7\%}$
\\
    &
    &
    $pp \rightarrow B^{0}_s$+$X$
    & $10.5 \pm 0.2\text{\scriptsize (stat)} \pm 0.8\text{\scriptsize (syst)} \pm 1.0\text{\scriptsize (bf)}$
    & $9.32(6)_{-11.5\%}^{+13.6\%}$
\\
    \hline
    \multirow{2}{*}{\setuplhcbtwo{} \cite{LHCb:2017vec}}
    &
    7\,TeV 
    & $pp\rightarrow B^{\pm}$+$X$
    & $43.0 \pm 0.2\text{\scriptsize (syst)} \pm 2.5\text{\scriptsize (stat)} \pm 1.7\text{\scriptsize (bf)}$
    & $42.2(1)_{-11.4\%}^{+13.9\%}$
    \\
    &
    13\,TeV 
    & $pp\rightarrow B^{\pm}$+$X$
    & $86.6 \pm 0.5\text{\scriptsize (stat)} \pm 5.4\text{\scriptsize (syst)} \pm 3.4\text{\scriptsize (bf)}$
    & $78.5(3)_{-9.3\%}^{+9.0\%}$
    \\ 
    \hline
    \multirow{2}{*}{\setuplhcbthree{} \cite{LHCb:2016qpe}}
    &
    7\,TeV
    &
    $pp \rightarrow B$+$X$
    & $72.0 \pm 0.3\text{\scriptsize (stat)} \pm 6.8\text{\scriptsize (syst)}$
    & $65.3(1)_{-10.5\%}^{+12.6\%}$
\\
    &
     13\,TeV
    &
    $pp \rightarrow B$+$X$
    & $144 \pm 1\text{\scriptsize (stat)} \pm 21\text{\scriptsize (syst)}$
    & $116.2(3)_{-12.3\%}^{+7.6\%}$
\\
    \hline
    \end{tabular}%
}%
\vspace{0.5cm}
    \caption{\label{tab:cs} Fiducial cross sections for the production of different $B$ hadron final states for various LHC analyses and compared against \minnlo{} predictions. See text for details.}
\end{center}
\end{table}

First, we consider various cross-section measurements of $B$ meson (and hadron) production 
at different LHC energies and by different experiments in \tab{tab:cs}, with standard experimental 
uncertainties (statistical, systematical, luminosity) and one related to the assumed 
branching fractions (bf) of the $B$ hadrons. 
These analyses measure the cross section for the production of different $B$ hadrons.
The \setupatlas{} 7\,TeV analysis of \citere{ATLAS:2013cia} and the \setupcms{} 13\,TeV analysis 
of \citere{CMS:2016plw} select only $B^+$ mesons, for instance. The \setuplhcbone{} analysis at 7\,TeV \cite{LHCb:2013vjr},
on the other hand, measures separately $B^\pm$, $B^0$ and $B^0_s$  meson cross sections (the latter two include also the charge conjugate mesons),
while the \setuplhcbtwo{} analysis \cite{LHCb:2017vec} includes both $B^+$ and  $B^-$ meson,
providing their cross sections at 7\,TeV and 13\,TeV.
Only the \setuplhcbthree{} study of \citere{LHCb:2016qpe} accounts for all relevant $B$ hadrons, including also 
some baryons, in order to provide predictions as close as possible to the originally produced bottom quarks.
By and large, it is quite remarkable how well our \minnlo{} predictions agree with the measured cross sections
within the quoted experimental and theoretical uncertainties, except for the \setupcms{} measurement at 13\,TeV,
where the \minnlo{} prediction is somewhat lower.
Moreover, the experimental and theoretical uncertainties are largely of similar size, which shows that 
NNLO+PS accuracy is required, also in view of future measurements.

\begin{figure}[p]
\begin{center}
\includegraphics[width=.4\textwidth, page=6]{plots/ATLAS.pdf}\hspace{1cm}
\includegraphics[width=.4\textwidth, page=3]{plots/ATLAS.pdf}
\begin{tabular}{cccc}
\hspace{-1.1cm}
\hstretch{1}{\includegraphics[width=.4\textwidth, page=2]{plots/ATLAS.pdf}}
&\hspace{-3.3cm}
\hstretch{1}{\includegraphics[width=.4\textwidth, page=1]{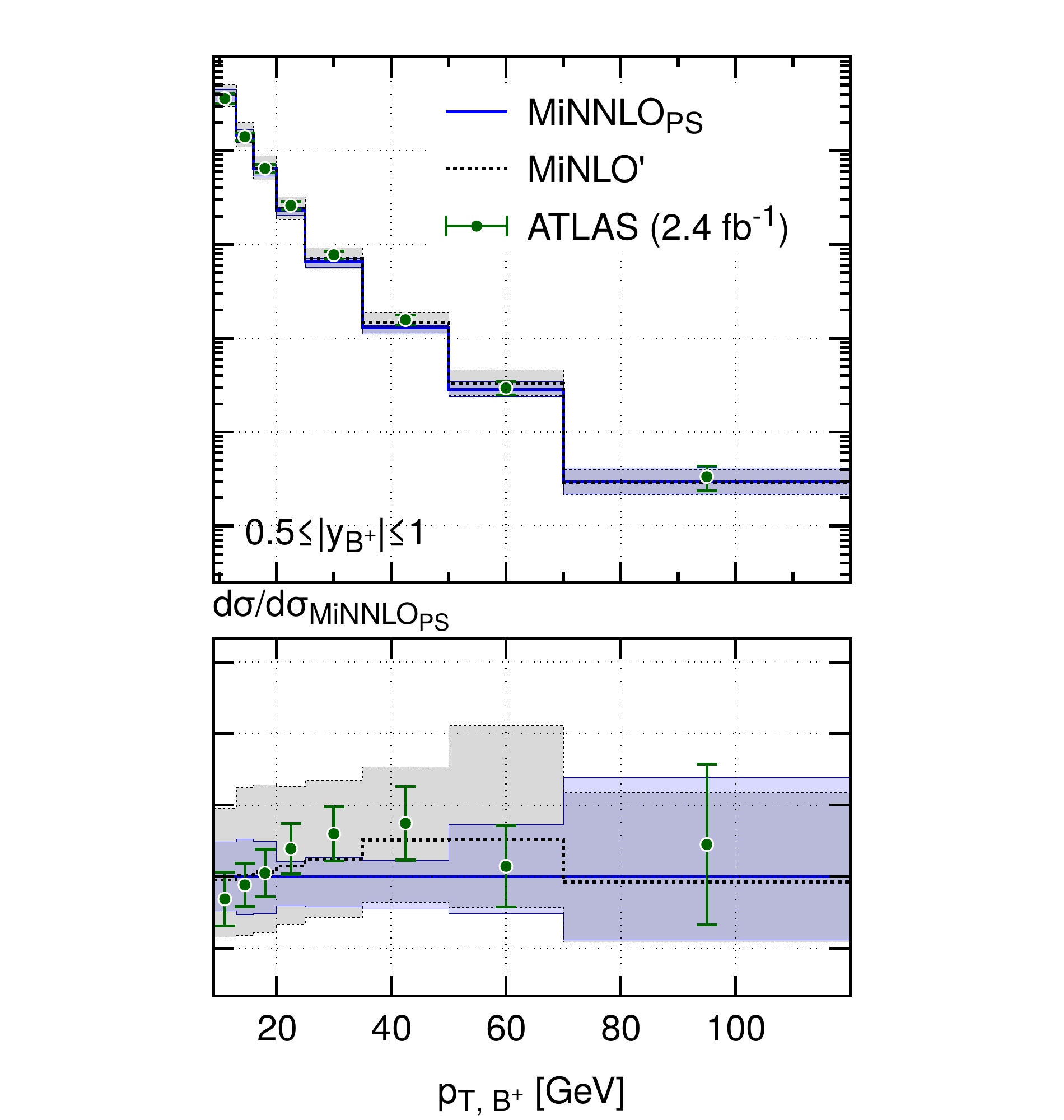}}
&\hspace{-3.3cm}
\hstretch{1}{\includegraphics[width=.4\textwidth, page=5]{plots/ATLAS.pdf}}
&\hspace{-3.3cm}
\hstretch{1}{\includegraphics[width=.4\textwidth, page=4]{plots/ATLAS.pdf}}
\hspace{-0.3cm}
\end{tabular}
\caption{\label{fig:ptb7atlas} Comparison to ATLAS 7\,TeV data \cite{ATLAS:2013cia}. See text for details.}
\end{center}
\end{figure}

We continue by studying our \minnlo{} predictions in comparison to differential measurements.
In \fig{fig:ptb7atlas} we consider the data (green points) from the $B^+$-meson analysis by \setupatlas{} 
at 7\,TeV for the $B^+$ rapidity ($y_{B^+}$) and transverse momentum ($p_{T,B^+}$). The upper
figures show the two single differential distributions in the selected phase space, while the lower
ones are the double differential distributions, namely the $p_{T,B^+}$ observable in slices of $y_{B^+}$.
From the $y_{B^+}$ distribution it is clear that NNLO corrections of the \minnlo{} prediction (blue, solid)
with respect to the \minlo{} result is completely flat and practically zero in this fiducial setup, while they 
induce a substantial reduction of the theoretical higher-order uncertainties estimated from scale variation. 
For the $p_{T,B^+}$ spectrum, on the other hand, we see that \minnlo{} predicts a softer 
behaviour in the tail of the distribution, which induces a slight improvement in the description of the data.
However, in either case, both \minlo{} and \minnlo{} predictions are in full agreement with the 
measured $y_{B^+}$ and $p_{T,B^+}$ distributions within uncertainties.
This is true, also for the double-differential $p_{T,B^+}$--$y_{B^+}$ results with the exception of a 
fluctuation of the data in a single bin. Overall, the picture remains the same though: \minnlo{} 
features much smaller uncertainty bands compared to \minlo{} , a softer $p_{T,B^+}$ spectrum in each $y_{B^+}$ slice,
and a remarkable agreement with data.

\begin{figure}[t!]
\begin{center}
\begin{tabular}{cccc}
\includegraphics[width=.38\textwidth, page=2]{plots/CMS_joined2.pdf}
&
\hspace{0.25cm}
\includegraphics[width=.38\textwidth, page=1]{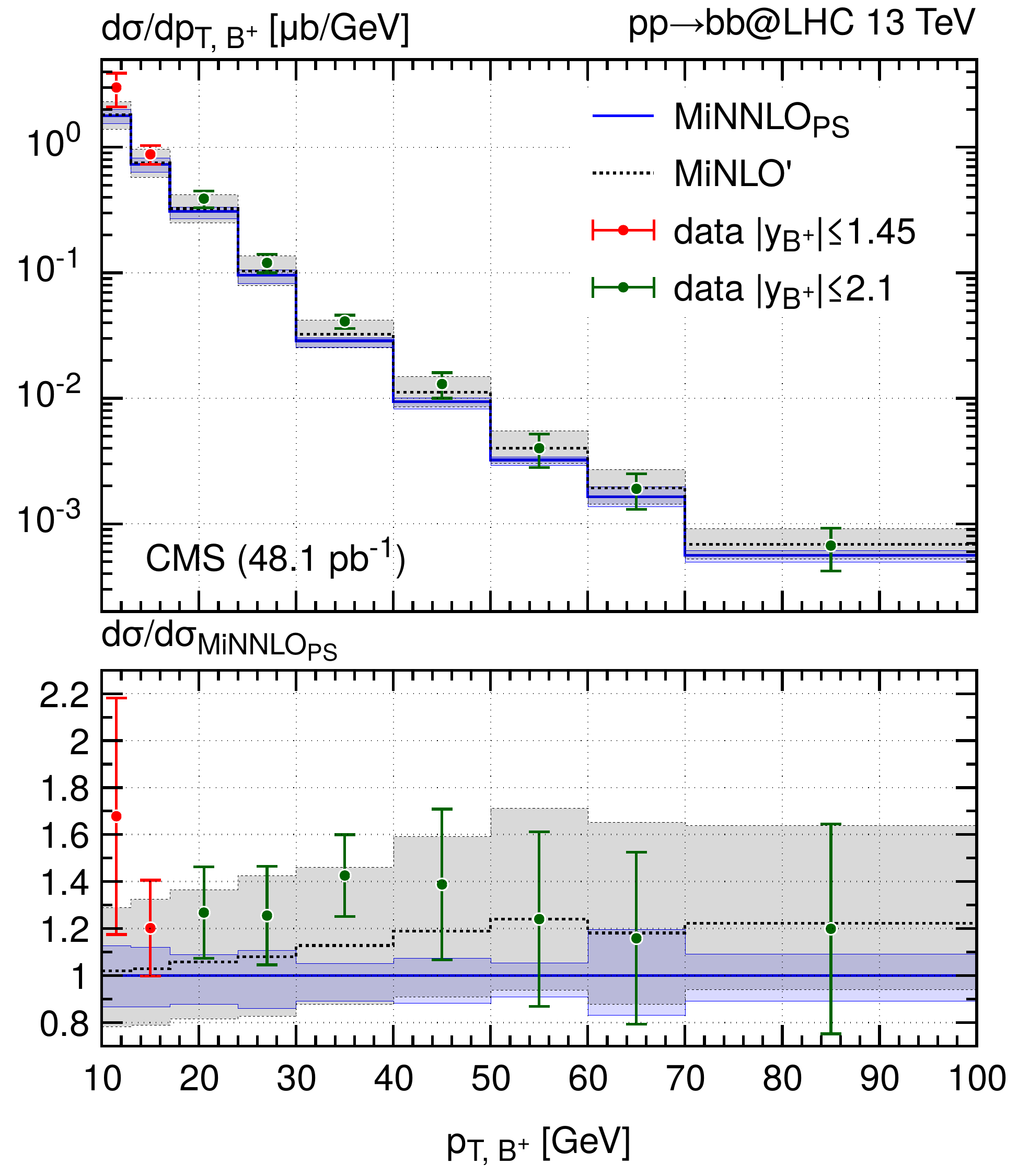}
\end{tabular}
\vspace*{1ex}
\caption{\label{fig:cms}  Comparison to CMS 13\,TeV data \cite{CMS:2016plw}. See text for details.}
\end{center}
\end{figure}

Next, we present a comparison against the \setupcms{} measurement at 13\,TeV of $y_{B^+}$ and $p_{T,B^+}$ in \fig{fig:cms}. 
In this analysis, the $10\le p_{T,B^+}\le 17$\,GeV region is measured with a smaller rapidity range (up to $|y|\le 1.45$)
and the extended range ($p_{T,B^+}\le 100$\,GeV) is measured up to $|y|\le 2.1$, which explains the differently labelled data points.
As one can see, all data points are consistently above the \minnlo{} predictions (despite being largely within the quoted uncertainties).
We already observed this (small) discrepancy for the measured cross section in \tab{tab:cs}. The shape of the 
differential distributions, on the other hand, are well described by the \minnlo{} predictions.

%\afterpage{
\begin{figure}[p!]
\begin{center}
\begin{tabular}{cccc}
\hspace{-.5cm}
\includegraphics[width=.26\textwidth, page=7]{plots/LHCb2_7TeV.pdf}
&
\hspace{-0.6cm}
\includegraphics[width=.26\textwidth, page=7]{plots/LHCb2_13TeV.pdf}
&
\hspace{-0.6cm}
\includegraphics[width=.26\textwidth, page=3]{plots/LHCb2_7TeV.pdf}
&
\hspace{-0.6cm}
\includegraphics[width=.26\textwidth, page=3]{plots/LHCb2_13TeV.pdf}
\end{tabular}
\begin{tabular}{ccccc}
\hspace{-0.8cm}
\hstretch{1}{\includegraphics[width=.25\textwidth, page=2]{plots/LHCb2_7TeV.pdf}}
&\hspace{-1.32cm}
\hstretch{1}{\includegraphics[width=.25\textwidth, page=1]{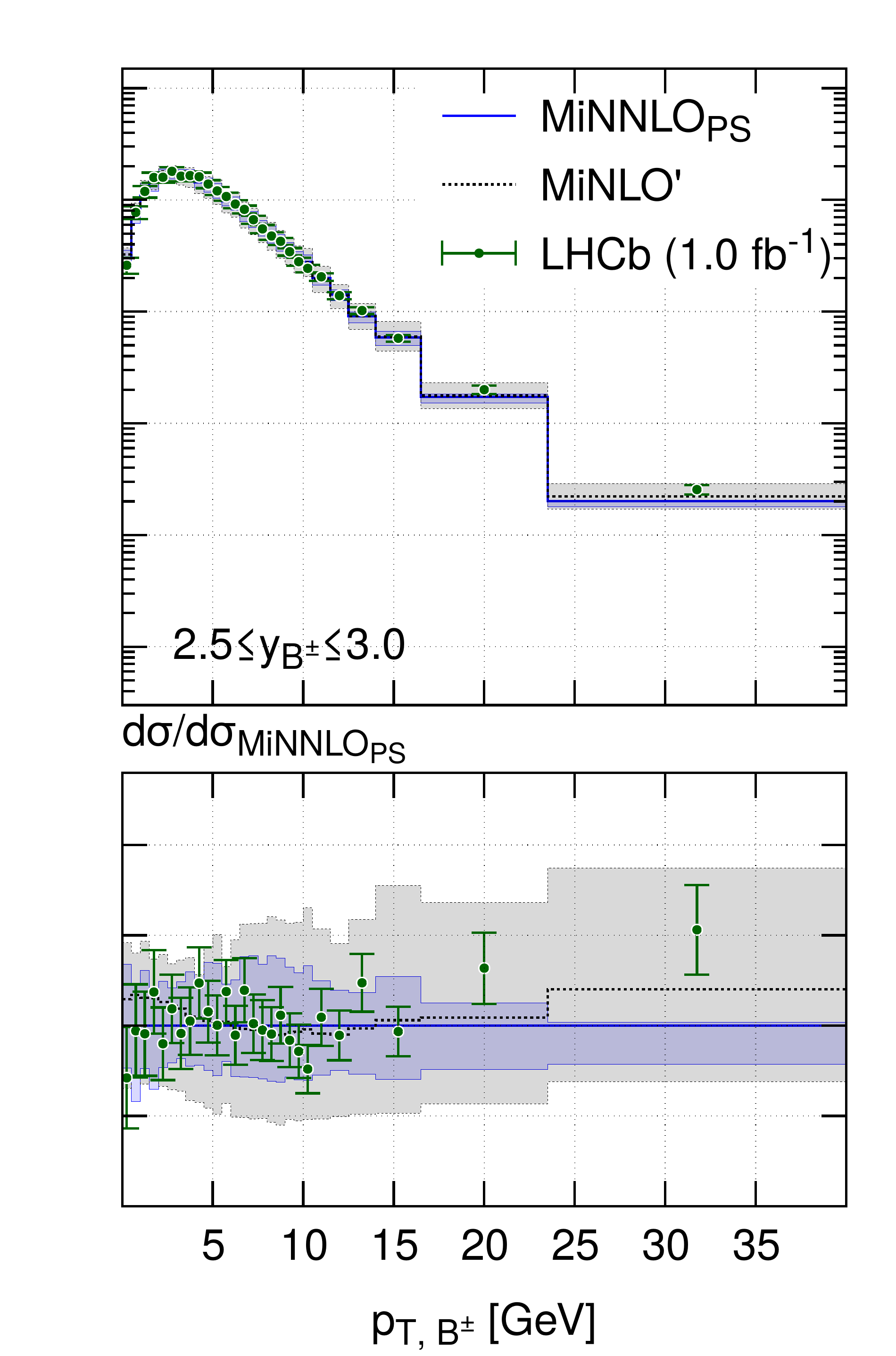}}
&\hspace{-1.32cm}
\hstretch{1}{\includegraphics[width=.25\textwidth, page=5]{plots/LHCb2_7TeV.pdf}}
&\hspace{-1.32cm}
\hstretch{1}{\includegraphics[width=.25\textwidth, page=4]{plots/LHCb2_7TeV.pdf}}
&\hspace{-1.32cm}
\hstretch{1}{\includegraphics[width=.25\textwidth, page=6]{plots/LHCb2_7TeV.pdf}}\\
\end{tabular}
%\vspace*{1ex}
%\caption{\label{fig:ptb7lhcb} 7 TeV LHCb pTB from \cite{LHCb:2017vec}. See text for details.}
%\end{center}
%%\end{figure}
%%\begin{figure}[t!]
%\begin{center}
\begin{tabular}{ccccc}
\hspace{-0.8cm}
\hstretch{1}{\includegraphics[width=.25\textwidth, page=2]{plots/LHCb2_13TeV.pdf}}
&\hspace{-1.32cm}
\hstretch{1}{\includegraphics[width=.25\textwidth, page=1]{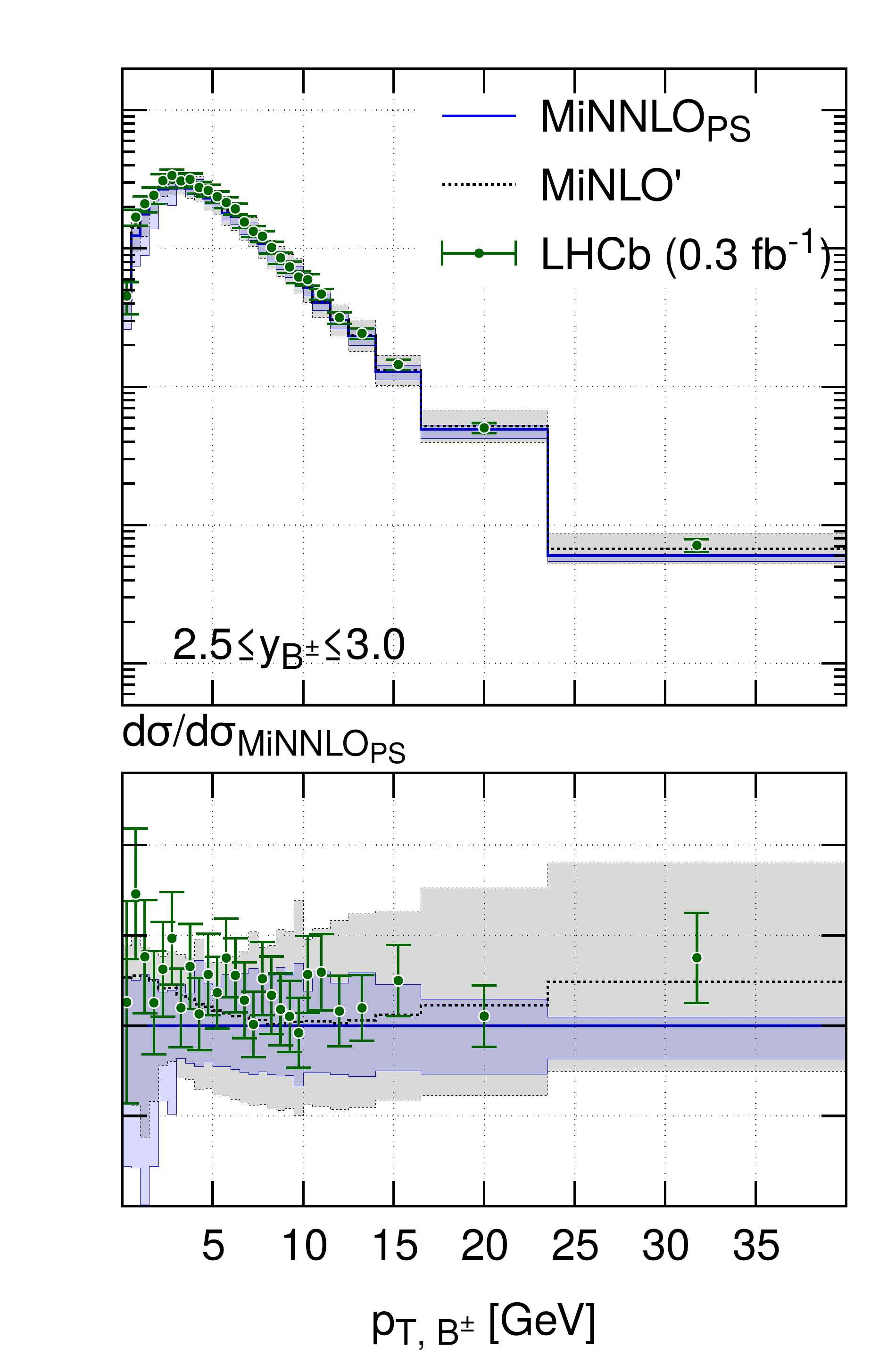}}
&\hspace{-1.32cm}
\hstretch{1}{\includegraphics[width=.25\textwidth, page=5]{plots/LHCb2_13TeV.pdf}}
&\hspace{-1.32cm}
\hstretch{1}{\includegraphics[width=.25\textwidth, page=4]{plots/LHCb2_13TeV.pdf}}
&\hspace{-1.32cm}
\hstretch{1}{\includegraphics[width=.25\textwidth, page=6]{plots/LHCb2_13TeV.pdf}}\\
\end{tabular}
\vspace*{1ex}
\caption{\label{fig:ptb713lhcb} Comparison to LHCb data at 7 and 13\,TeV \cite{LHCb:2017vec}. See text for details.
}
\end{center}
\end{figure}
%\clearpage
%}

A very similar picture as for the ATLAS 7\,TeV comparison in \fig{fig:ptb7atlas}
emerges for the \setuplhcbtwo{} $B^\pm$ data at 7\,TeV and 13\,TeV in \fig{fig:ptb713lhcb},
which shows the rapidity ($y_{B^\pm}$) and transverse momentum ($p_{T,B^\pm}$) of the $B^\pm$ mesons.
One should notice that due to their asymmetric detector design LHCb can measure only in one rapidity direction, 
 but up to significantly larger values of it. Moreover, the \setuplhcbtwo{} exhibits a very fine binning in $p_{T,B^\pm}$.
Once again, we observe an extremely good agreement with the \minnlo{} predictions. This is is true not only 
in terms of normalization, but also in terms of the shapes, especially for the finely binned  $p_{T,B^\pm}$ distribution.

\begin{figure}[t!]
\begin{center}
\begin{tabular}{cc}
\includegraphics[width=.38\textwidth, page=2]{plots/LHCb3.pdf}
&
\hspace{0.25cm}
\includegraphics[width=.38\textwidth, page=1]{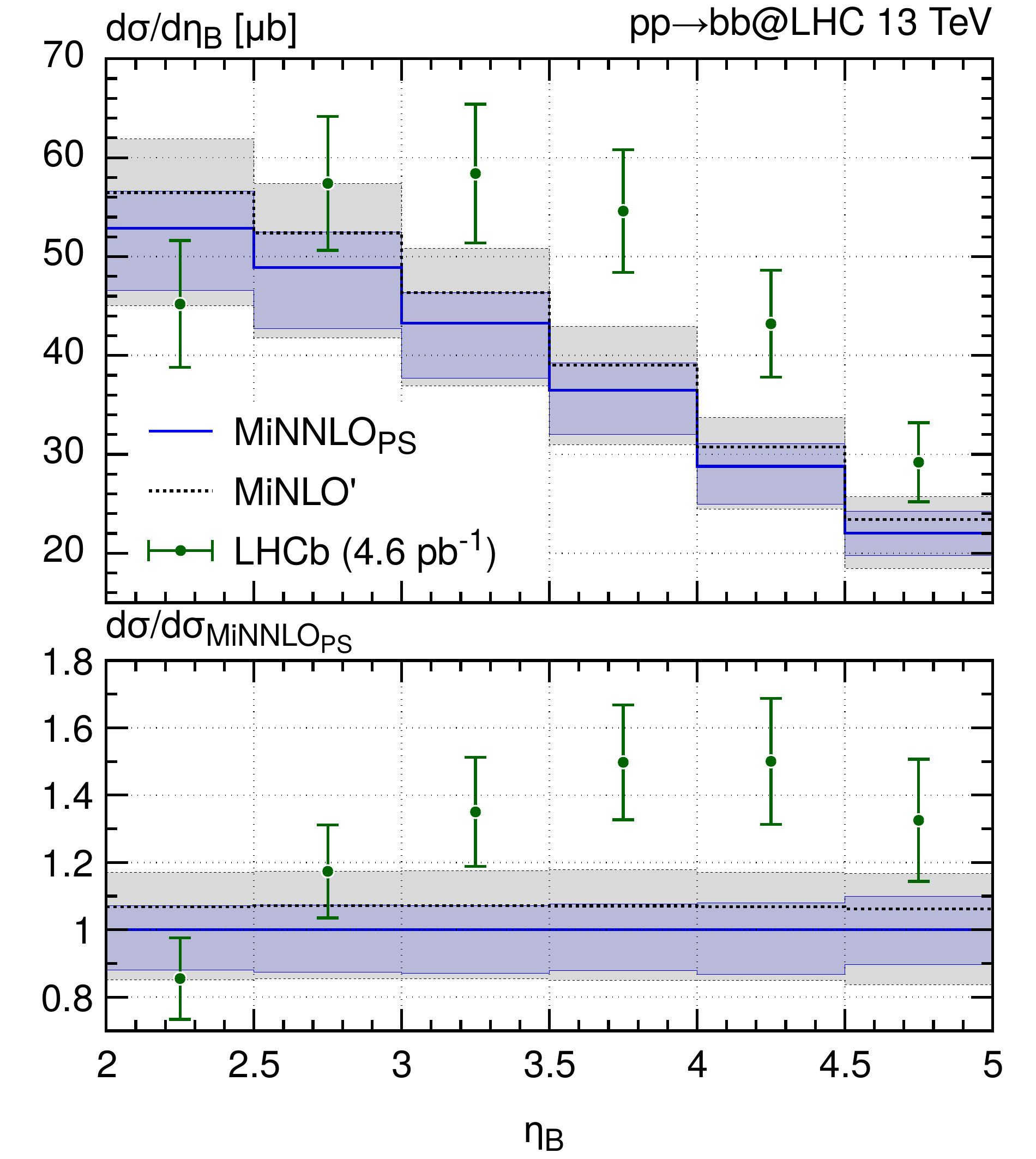}%\\
%\includegraphics[width=.34\textwidth, page=14]{plots/LHCb3.pdf}
%&
%\hspace{-0.55cm}
%\includegraphics[width=.34\textwidth]{plots/LHCb@13TeV_y_Bplus_Bminus.pdf}
\end{tabular}
\vspace*{1ex}
\caption{\label{fig:rapidity713lhcb} Comparison to LHCb rapidity distributions at 7 and 13\,TeV \cite{LHCb:2016qpe}. See text for details.}
\end{center}
\end{figure}

In \fig{fig:rapidity713lhcb} we show results for the \setuplhcbthree{} analysis for the pseudorapidity of the $B$ hadron ($\eta_B$), which
includes all relevant $B$ hadrons that yield a sufficiently large contribution to the cross section. In this case, the $\eta_B$ shape, especially of the 13\,TeV
measurement, cannot be reproduced by our predictions. This is in line with the fact that neither the FONLL result quoted in this analysis~\cite{LHCb:2016qpe} nor the recent NNLO QCD calculation \cite{Catani:2020kkl} predict such a shape. Given that all other rapidity measurements are in excellent agreement
with SM predictions, especially those presented here from our \minnlo{} generator, it seems unlikely that this is induced by some new-physics phenomenon.

\begin{figure}[t!]
\begin{center}
\begin{tabular}{ccc}
\hspace{-1cm}
\includegraphics[width=.37\textwidth, page=2]{plots/ratios.pdf}
&
\hspace{-0.8cm}
\includegraphics[width=.37\textwidth, page=3]{plots/ratios.pdf}
&
\hspace{-0.8cm}
\includegraphics[width=.37\textwidth, page=1]{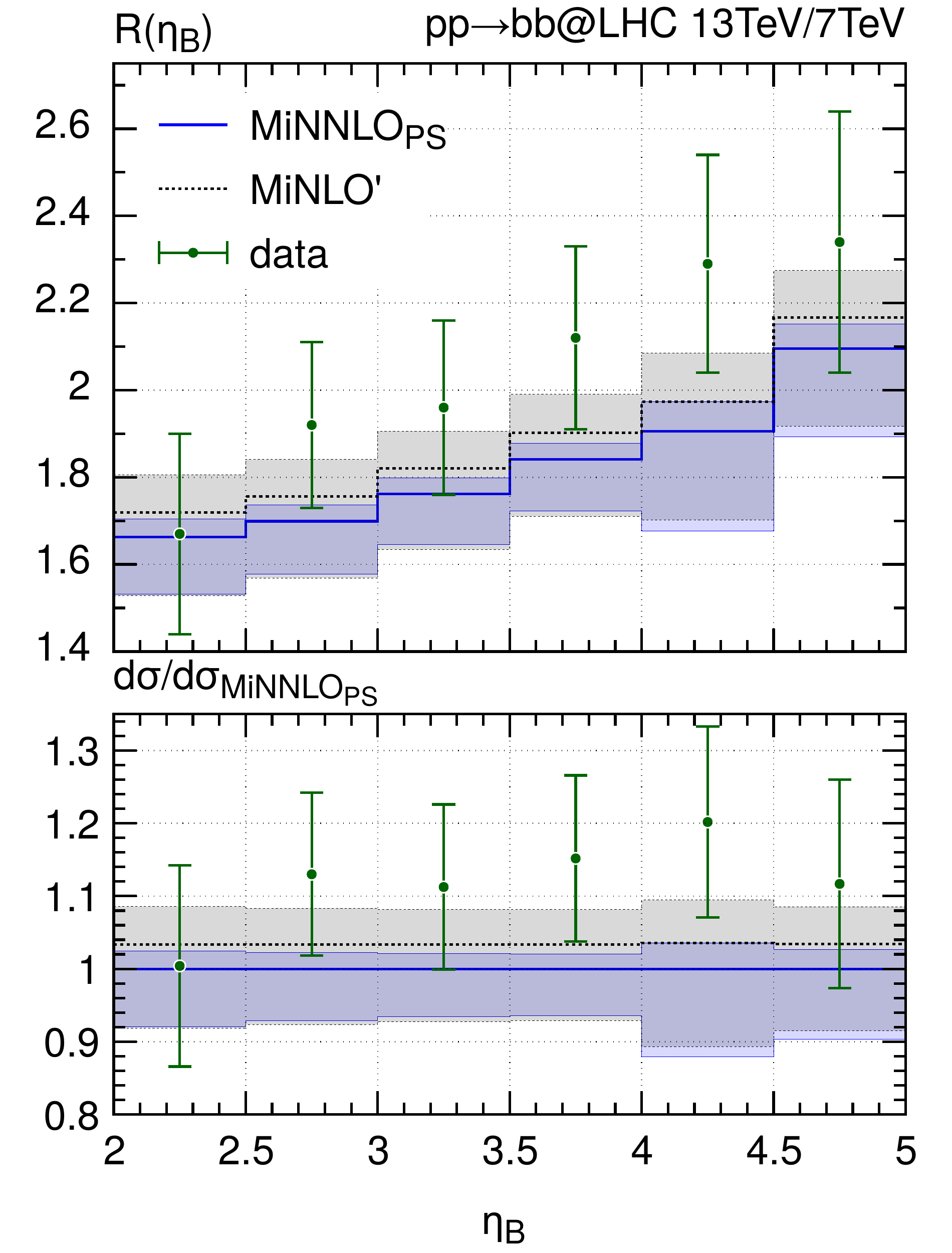}
\end{tabular}
\vspace*{1ex}
\caption{\label{fig:ratios} Comparison to 13\,TeV/7\,TeV ratios measured by LHCb \cite{LHCb:2017vec,LHCb:2016qpe}. See text for details.}
\end{center}
\end{figure}

Finally, we briefly comment on the 13\,TeV/7\,TeV ratios presented in the \setuplhcbtwo{} analysis for the $p_{T,B^\pm}$ and $y_{B^\pm}$ distributions, and in the
 \setuplhcbthree{} analysis for the
$\eta_{B}$ distribution, shown in \fig{fig:ratios}. In all cases, the \minnlo{} corrections are quite small ($\sim -5\%$) and flat in each distribution. Moreover, 
the scale uncertainties are reduced to the $\sim 10\%$ level and slightly smaller for \minnlo{} compared to \minlo{}. The experimental data is full agreement with the 
predictions.

%%%%%%%%%%%%%%%%%%%%%%%%%%%%%%

\section{Summary}

To summarize, we have presented a fully exclusive simulation of $B$
hadron production at the LHC, which describes the underlying hard
process $pp\to b\bar{b}$ at NNLO QCD accuracy. We have validated the
accuracy of our NNLO+PS calculation for bottom-quark pair production
against fixed-order NNLO QCD predictions.  The comparison to LHC data
from various analyses by ATLAS, CMS and LHCb at 7 and/or 13\,TeV shows
that the NNLO QCD corrections are important to reach an accurate
description of $B$ meson (and $B$ hadron) observables. Not only do
we find very good agreement of our \minnlo{} predictions with data
both at the cross-section and at the distribution level, we also
observe a clear reduction of uncertainties with respect to lower order
predictions.  We reckon that our new $b\bar{b}$ \minnlo{} generator,
which can be used to simulate fully exclusively the kinematics of the
bottom-flavoured hadronic final states, will be particularly useful
for future $B$ hadron measurements at the LHC.
The code will be made publicly available within the \POWHEGBOXRES{}
framework.

Our calculation allows us to have an accurate and realistic description 
of $b$-jet cross sections as well, enabling a direct comparison to 
$b$-jet measurements at the LHC. In this context, also the impact of different algorithms 
to define the jet flavour  \cite{Banfi:2006hf,Buckley:2015gua,Ilten:2017rbd,Caletti:2021oor,Fedkevych:2022mid,Caletti:2022hnc,Caletti:2022glq,Czakon:2022wam,Gauld:2022lem} 
can be studied, which 
received quite some attention recently.
We leave such studies to future work.

\noindent {\bf Acknowledgements.}
We would like to thank Pier Francesco Monni, Paolo Nason and Emanuele Re for comments on the manuscript.
We have used the Max Planck Computing and Data Facility (MPCDF) in
Garching to carry out all simulations presented here.

\setlength{\bibsep}{3.1pt}
\renewcommand{\em}{}
\bibliographystyle{apsrev4-1}
\bibliography{MiNNLO}

\begin{thebibliography}{91}%
\makeatletter
\providecommand \@ifxundefined [1]{%
 \@ifx{#1\undefined}
}%
\providecommand \@ifnum [1]{%
 \ifnum #1\expandafter \@firstoftwo
 \else \expandafter \@secondoftwo
 \fi
}%
\providecommand \@ifx [1]{%
 \ifx #1\expandafter \@firstoftwo
 \else \expandafter \@secondoftwo
 \fi
}%
\providecommand \natexlab [1]{#1}%
\providecommand \enquote  [1]{``#1''}%
\providecommand \bibnamefont  [1]{#1}%
\providecommand \bibfnamefont [1]{#1}%
\providecommand \citenamefont [1]{#1}%
\providecommand \href@noop [0]{\@secondoftwo}%
\providecommand \href [0]{\begingroup \@sanitize@url \@href}%
\providecommand \@href[1]{\@@startlink{#1}\@@href}%
\providecommand \@@href[1]{\endgroup#1\@@endlink}%
\providecommand \@sanitize@url [0]{\catcode `\\12\catcode `\$12\catcode
  `\&12\catcode `\#12\catcode `\^12\catcode `\_12\catcode `\%12\relax}%
\providecommand \@@startlink[1]{}%
\providecommand \@@endlink[0]{}%
\providecommand \url  [0]{\begingroup\@sanitize@url \@url }%
\providecommand \@url [1]{\endgroup\@href {#1}{\urlprefix }}%
\providecommand \urlprefix  [0]{URL }%
\providecommand \Eprint [0]{\href }%
\providecommand \doibase [0]{http://dx.doi.org/}%
\providecommand \selectlanguage [0]{\@gobble}%
\providecommand \bibinfo  [0]{\@secondoftwo}%
\providecommand \bibfield  [0]{\@secondoftwo}%
\providecommand \translation [1]{[#1]}%
\providecommand \BibitemOpen [0]{}%
\providecommand \bibitemStop [0]{}%
\providecommand \bibitemNoStop [0]{.\EOS\space}%
\providecommand \EOS [0]{\spacefactor3000\relax}%
\providecommand \BibitemShut  [1]{\csname bibitem#1\endcsname}%
\let\auto@bib@innerbib\@empty
%</preamble>
\bibitem [{\citenamefont {Albajar}\ \emph {et~al.}(1987)\citenamefont {Albajar}
  \emph {et~al.}}]{Albajar:1986iu}%
  \BibitemOpen
  \bibfield  {author} {\bibinfo {author} {\bibfnamefont {C.}~\bibnamefont
  {Albajar}} \emph {et~al.} (\bibinfo {collaboration} {UA1}),\ }\href {\doibase
  10.1016/0370-2693(87)90287-5} {\bibfield  {journal} {\bibinfo  {journal}
  {Phys. Lett. B}\ }\textbf {\bibinfo {volume} {186}},\ \bibinfo {pages} {237}
  (\bibinfo {year} {1987})}\BibitemShut {NoStop}%
\bibitem [{\citenamefont {Albajar}\ \emph {et~al.}(1988)\citenamefont {Albajar}
  \emph {et~al.}}]{Albajar:1988th}%
  \BibitemOpen
  \bibfield  {author} {\bibinfo {author} {\bibfnamefont {C.}~\bibnamefont
  {Albajar}} \emph {et~al.} (\bibinfo {collaboration} {UA1}),\ }\href {\doibase
  10.1016/0370-2693(88)91785-6} {\bibfield  {journal} {\bibinfo  {journal}
  {Phys. Lett. B}\ }\textbf {\bibinfo {volume} {213}},\ \bibinfo {pages} {405}
  (\bibinfo {year} {1988})}\BibitemShut {NoStop}%
\bibitem [{\citenamefont {Abe}\ \emph {et~al.}(1995)\citenamefont {Abe} \emph
  {et~al.}}]{Abe:1995dv}%
  \BibitemOpen
  \bibfield  {author} {\bibinfo {author} {\bibfnamefont {F.}~\bibnamefont
  {Abe}} \emph {et~al.} (\bibinfo {collaboration} {CDF}),\ }\href {\doibase
  10.1103/PhysRevLett.75.1451} {\bibfield  {journal} {\bibinfo  {journal}
  {Phys. Rev. Lett.}\ }\textbf {\bibinfo {volume} {75}},\ \bibinfo {pages}
  {1451} (\bibinfo {year} {1995})},\ \Eprint
  {http://arxiv.org/abs/hep-ex/9503013} {arXiv:hep-ex/9503013}\BibitemShut
  {NoStop}%
\bibitem [{\citenamefont {Acosta}\ \emph {et~al.}(2002)\citenamefont {Acosta}
  \emph {et~al.}}]{Acosta:2001rz}%
  \BibitemOpen
  \bibfield  {author} {\bibinfo {author} {\bibfnamefont {D.}~\bibnamefont
  {Acosta}} \emph {et~al.} (\bibinfo {collaboration} {CDF}),\ }\href {\doibase
  10.1103/PhysRevD.65.052005} {\bibfield  {journal} {\bibinfo  {journal} {Phys.
  Rev. D}\ }\textbf {\bibinfo {volume} {65}},\ \bibinfo {pages} {052005}
  (\bibinfo {year} {2002})},\ \Eprint {http://arxiv.org/abs/hep-ph/0111359}
  {arXiv:hep-ph/0111359}\BibitemShut {NoStop}%
\bibitem [{\citenamefont {Acosta}\ \emph {et~al.}(2005)\citenamefont {Acosta}
  \emph {et~al.}}]{Acosta:2004yw}%
  \BibitemOpen
  \bibfield  {author} {\bibinfo {author} {\bibfnamefont {D.}~\bibnamefont
  {Acosta}} \emph {et~al.} (\bibinfo {collaboration} {CDF}),\ }\href {\doibase
  10.1103/PhysRevD.71.032001} {\bibfield  {journal} {\bibinfo  {journal} {Phys.
  Rev. D}\ }\textbf {\bibinfo {volume} {71}},\ \bibinfo {pages} {032001}
  (\bibinfo {year} {2005})},\ \Eprint {http://arxiv.org/abs/hep-ex/0412071}
  {arXiv:hep-ex/0412071}\BibitemShut {NoStop}%
\bibitem [{\citenamefont {Abulencia}\ \emph {et~al.}(2007)\citenamefont
  {Abulencia} \emph {et~al.}}]{Abulencia:2006ps}%
  \BibitemOpen
  \bibfield  {author} {\bibinfo {author} {\bibfnamefont {A.}~\bibnamefont
  {Abulencia}} \emph {et~al.} (\bibinfo {collaboration} {CDF}),\ }\href
  {\doibase 10.1103/PhysRevD.75.012010} {\bibfield  {journal} {\bibinfo
  {journal} {Phys. Rev. D}\ }\textbf {\bibinfo {volume} {75}},\ \bibinfo
  {pages} {012010} (\bibinfo {year} {2007})},\ \Eprint
  {http://arxiv.org/abs/hep-ex/0612015} {arXiv:hep-ex/0612015}\BibitemShut
  {NoStop}%
\bibitem [{\citenamefont {Abachi}\ \emph {et~al.}(1995)\citenamefont {Abachi}
  \emph {et~al.}}]{Abachi:1994kj}%
  \BibitemOpen
  \bibfield  {author} {\bibinfo {author} {\bibfnamefont {S.}~\bibnamefont
  {Abachi}} \emph {et~al.} (\bibinfo {collaboration} {D0}),\ }\href {\doibase
  10.1103/PhysRevLett.74.3548} {\bibfield  {journal} {\bibinfo  {journal}
  {Phys. Rev. Lett.}\ }\textbf {\bibinfo {volume} {74}},\ \bibinfo {pages}
  {3548} (\bibinfo {year} {1995})}\BibitemShut {NoStop}%
\bibitem [{\citenamefont {Abbott}\ \emph {et~al.}(2000)\citenamefont {Abbott}
  \emph {et~al.}}]{Abbott:1999se}%
  \BibitemOpen
  \bibfield  {author} {\bibinfo {author} {\bibfnamefont {B.}~\bibnamefont
  {Abbott}} \emph {et~al.} (\bibinfo {collaboration} {D0}),\ }\href {\doibase
  10.1016/S0370-2693(00)00844-3} {\bibfield  {journal} {\bibinfo  {journal}
  {Phys. Lett. B}\ }\textbf {\bibinfo {volume} {487}},\ \bibinfo {pages} {264}
  (\bibinfo {year} {2000})},\ \Eprint {http://arxiv.org/abs/hep-ex/9905024}
  {arXiv:hep-ex/9905024}\BibitemShut {NoStop}%
\bibitem [{\citenamefont {Abelev}\ \emph {et~al.}(2014)\citenamefont {Abelev}
  \emph {et~al.}}]{Abelev:2014hla}%
  \BibitemOpen
  \bibfield  {author} {\bibinfo {author} {\bibfnamefont {B.~B.}\ \bibnamefont
  {Abelev}} \emph {et~al.} (\bibinfo {collaboration} {ALICE}),\ }\href
  {\doibase 10.1016/j.physletb.2014.09.026} {\bibfield  {journal} {\bibinfo
  {journal} {Phys. Lett. B}\ }\textbf {\bibinfo {volume} {738}},\ \bibinfo
  {pages} {97} (\bibinfo {year} {2014})},\ \Eprint
  {http://arxiv.org/abs/1405.4144} {arXiv:1405.4144 [nucl-ex]}\BibitemShut
  {NoStop}%
\bibitem [{\citenamefont {Abelev}\ \emph {et~al.}(2013)\citenamefont {Abelev}
  \emph {et~al.}}]{Abelev:2012sca}%
  \BibitemOpen
  \bibfield  {author} {\bibinfo {author} {\bibfnamefont {B.}~\bibnamefont
  {Abelev}} \emph {et~al.} (\bibinfo {collaboration} {ALICE}),\ }\href
  {\doibase 10.1016/j.physletb.2013.01.069} {\bibfield  {journal} {\bibinfo
  {journal} {Phys. Lett. B}\ }\textbf {\bibinfo {volume} {721}},\ \bibinfo
  {pages} {13} (\bibinfo {year} {2013})},\ \bibinfo {note} {[Erratum:
  Phys.Lett.B 763, 507--509 (2016)]},\ \Eprint {http://arxiv.org/abs/1208.1902}
  {arXiv:1208.1902 [hep-ex]}\BibitemShut {NoStop}%
\bibitem [{\citenamefont {Aad}\ \emph {et~al.}(2012)\citenamefont {Aad} \emph
  {et~al.}}]{Aad:2012jga}%
  \BibitemOpen
  \bibfield  {author} {\bibinfo {author} {\bibfnamefont {G.}~\bibnamefont
  {Aad}} \emph {et~al.} (\bibinfo {collaboration} {ATLAS}),\ }\href {\doibase
  10.1016/j.nuclphysb.2012.07.009} {\bibfield  {journal} {\bibinfo  {journal}
  {Nucl. Phys. B}\ }\textbf {\bibinfo {volume} {864}},\ \bibinfo {pages} {341}
  (\bibinfo {year} {2012})},\ \Eprint {http://arxiv.org/abs/1206.3122}
  {arXiv:1206.3122 [hep-ex]}\BibitemShut {NoStop}%
\bibitem [{\citenamefont {Aad}\ \emph {et~al.}(2013)\citenamefont {Aad} \emph
  {et~al.}}]{ATLAS:2013cia}%
  \BibitemOpen
  \bibfield  {author} {\bibinfo {author} {\bibfnamefont {G.}~\bibnamefont
  {Aad}} \emph {et~al.} (\bibinfo {collaboration} {ATLAS}),\ }\href {\doibase
  10.1007/JHEP10(2013)042} {\bibfield  {journal} {\bibinfo  {journal} {JHEP}\
  }\textbf {\bibinfo {volume} {10}},\ \bibinfo {pages} {042} (\bibinfo {year}
  {2013})},\ \Eprint {http://arxiv.org/abs/1307.0126} {arXiv:1307.0126
  [hep-ex]}\BibitemShut {NoStop}%
\bibitem [{\citenamefont {Khachatryan}\ \emph {et~al.}(2011)\citenamefont
  {Khachatryan} \emph {et~al.}}]{Khachatryan:2011mk}%
  \BibitemOpen
  \bibfield  {author} {\bibinfo {author} {\bibfnamefont {V.}~\bibnamefont
  {Khachatryan}} \emph {et~al.} (\bibinfo {collaboration} {CMS}),\ }\href
  {\doibase 10.1103/PhysRevLett.106.112001} {\bibfield  {journal} {\bibinfo
  {journal} {Phys. Rev. Lett.}\ }\textbf {\bibinfo {volume} {106}},\ \bibinfo
  {pages} {112001} (\bibinfo {year} {2011})},\ \Eprint
  {http://arxiv.org/abs/1101.0131} {arXiv:1101.0131 [hep-ex]}\BibitemShut
  {NoStop}%
\bibitem [{\citenamefont {Chatrchyan}\ \emph {et~al.}(2011)\citenamefont
  {Chatrchyan} \emph {et~al.}}]{Chatrchyan:2011pw}%
  \BibitemOpen
  \bibfield  {author} {\bibinfo {author} {\bibfnamefont {S.}~\bibnamefont
  {Chatrchyan}} \emph {et~al.} (\bibinfo {collaboration} {CMS}),\ }\href
  {\doibase 10.1103/PhysRevLett.106.252001} {\bibfield  {journal} {\bibinfo
  {journal} {Phys. Rev. Lett.}\ }\textbf {\bibinfo {volume} {106}},\ \bibinfo
  {pages} {252001} (\bibinfo {year} {2011})},\ \Eprint
  {http://arxiv.org/abs/1104.2892} {arXiv:1104.2892 [hep-ex]}\BibitemShut
  {NoStop}%
\bibitem [{\citenamefont {Chatrchyan}\ \emph {et~al.}(2012)\citenamefont
  {Chatrchyan} \emph {et~al.}}]{Chatrchyan:2012hw}%
  \BibitemOpen
  \bibfield  {author} {\bibinfo {author} {\bibfnamefont {S.}~\bibnamefont
  {Chatrchyan}} \emph {et~al.} (\bibinfo {collaboration} {CMS}),\ }\href
  {\doibase 10.1007/JHEP06(2012)110} {\bibfield  {journal} {\bibinfo  {journal}
  {JHEP}\ }\textbf {\bibinfo {volume} {06}},\ \bibinfo {pages} {110} (\bibinfo
  {year} {2012})},\ \Eprint {http://arxiv.org/abs/1203.3458} {arXiv:1203.3458
  [hep-ex]}\BibitemShut {NoStop}%
\bibitem [{\citenamefont {Khachatryan}\ \emph {et~al.}(2017)\citenamefont
  {Khachatryan} \emph {et~al.}}]{CMS:2016plw}%
  \BibitemOpen
  \bibfield  {author} {\bibinfo {author} {\bibfnamefont {V.}~\bibnamefont
  {Khachatryan}} \emph {et~al.} (\bibinfo {collaboration} {CMS}),\ }\href
  {\doibase 10.1016/j.physletb.2017.05.074} {\bibfield  {journal} {\bibinfo
  {journal} {Phys. Lett. B}\ }\textbf {\bibinfo {volume} {771}},\ \bibinfo
  {pages} {435} (\bibinfo {year} {2017})},\ \Eprint
  {http://arxiv.org/abs/1609.00873} {arXiv:1609.00873 [hep-ex]}\BibitemShut
  {NoStop}%
\bibitem [{\citenamefont {Aaij}\ \emph {et~al.}(2010)\citenamefont {Aaij} \emph
  {et~al.}}]{Aaij:2010gn}%
  \BibitemOpen
  \bibfield  {author} {\bibinfo {author} {\bibfnamefont {R.}~\bibnamefont
  {Aaij}} \emph {et~al.} (\bibinfo {collaboration} {LHCb}),\ }\href {\doibase
  10.1016/j.physletb.2010.10.010} {\bibfield  {journal} {\bibinfo  {journal}
  {Phys. Lett. B}\ }\textbf {\bibinfo {volume} {694}},\ \bibinfo {pages} {209}
  (\bibinfo {year} {2010})},\ \Eprint {http://arxiv.org/abs/1009.2731}
  {arXiv:1009.2731 [hep-ex]}\BibitemShut {NoStop}%
\bibitem [{\citenamefont {Aaij}\ \emph {et~al.}(2012)\citenamefont {Aaij} \emph
  {et~al.}}]{Aaij:2012jd}%
  \BibitemOpen
  \bibfield  {author} {\bibinfo {author} {\bibfnamefont {R.}~\bibnamefont
  {Aaij}} \emph {et~al.} (\bibinfo {collaboration} {LHCb}),\ }\href {\doibase
  10.1007/JHEP04(2012)093} {\bibfield  {journal} {\bibinfo  {journal} {JHEP}\
  }\textbf {\bibinfo {volume} {04}},\ \bibinfo {pages} {093} (\bibinfo {year}
  {2012})},\ \Eprint {http://arxiv.org/abs/1202.4812} {arXiv:1202.4812
  [hep-ex]}\BibitemShut {NoStop}%
\bibitem [{\citenamefont {Aaij}\ \emph {et~al.}(2013)\citenamefont {Aaij} \emph
  {et~al.}}]{LHCb:2013vjr}%
  \BibitemOpen
  \bibfield  {author} {\bibinfo {author} {\bibfnamefont {R.}~\bibnamefont
  {Aaij}} \emph {et~al.} (\bibinfo {collaboration} {LHCb}),\ }\href {\doibase
  10.1007/JHEP08(2013)117} {\bibfield  {journal} {\bibinfo  {journal} {JHEP}\
  }\textbf {\bibinfo {volume} {08}},\ \bibinfo {pages} {117} (\bibinfo {year}
  {2013})},\ \Eprint {http://arxiv.org/abs/1306.3663} {arXiv:1306.3663
  [hep-ex]}\BibitemShut {NoStop}%
\bibitem [{\citenamefont {Aaij}\ \emph
  {et~al.}(2017{\natexlab{a}})\citenamefont {Aaij} \emph
  {et~al.}}]{LHCb:2016qpe}%
  \BibitemOpen
  \bibfield  {author} {\bibinfo {author} {\bibfnamefont {R.}~\bibnamefont
  {Aaij}} \emph {et~al.} (\bibinfo {collaboration} {LHCb}),\ }\href {\doibase
  10.1103/PhysRevLett.118.052002} {\bibfield  {journal} {\bibinfo  {journal}
  {Phys. Rev. Lett.}\ }\textbf {\bibinfo {volume} {118}},\ \bibinfo {pages}
  {052002} (\bibinfo {year} {2017}{\natexlab{a}})},\ \bibinfo {note} {[Erratum:
  Phys.Rev.Lett. 119, 169901 (2017)]},\ \Eprint
  {http://arxiv.org/abs/1612.05140} {arXiv:1612.05140 [hep-ex]}\BibitemShut
  {NoStop}%
\bibitem [{\citenamefont {Aaij}\ \emph
  {et~al.}(2017{\natexlab{b}})\citenamefont {Aaij} \emph
  {et~al.}}]{LHCb:2017vec}%
  \BibitemOpen
  \bibfield  {author} {\bibinfo {author} {\bibfnamefont {R.}~\bibnamefont
  {Aaij}} \emph {et~al.} (\bibinfo {collaboration} {LHCb}),\ }\href {\doibase
  10.1007/JHEP12(2017)026} {\bibfield  {journal} {\bibinfo  {journal} {JHEP}\
  }\textbf {\bibinfo {volume} {12}},\ \bibinfo {pages} {026} (\bibinfo {year}
  {2017}{\natexlab{b}})},\ \Eprint {http://arxiv.org/abs/1710.04921}
  {arXiv:1710.04921 [hep-ex]}\BibitemShut {NoStop}%
\bibitem [{\citenamefont {Nason}\ \emph {et~al.}(1988)\citenamefont {Nason},
  \citenamefont {Dawson},\ and\ \citenamefont {Ellis}}]{Nason:1987xz}%
  \BibitemOpen
  \bibfield  {author} {\bibinfo {author} {\bibfnamefont {P.}~\bibnamefont
  {Nason}}, \bibinfo {author} {\bibfnamefont {S.}~\bibnamefont {Dawson}}\ and\
  \bibinfo {author} {\bibfnamefont {R.~K.}\ \bibnamefont {Ellis}},\ }\href
  {\doibase 10.1016/0550-3213(88)90422-1} {\bibfield  {journal} {\bibinfo
  {journal} {Nucl. Phys. B}\ }\textbf {\bibinfo {volume} {303}},\ \bibinfo
  {pages} {607} (\bibinfo {year} {1988})}\BibitemShut {NoStop}%
\bibitem [{\citenamefont {Nason}\ \emph {et~al.}(1989)\citenamefont {Nason},
  \citenamefont {Dawson},\ and\ \citenamefont {Ellis}}]{Nason:1989zy}%
  \BibitemOpen
  \bibfield  {author} {\bibinfo {author} {\bibfnamefont {P.}~\bibnamefont
  {Nason}}, \bibinfo {author} {\bibfnamefont {S.}~\bibnamefont {Dawson}}\ and\
  \bibinfo {author} {\bibfnamefont {R.~K.}\ \bibnamefont {Ellis}},\ }\href
  {\doibase 10.1016/0550-3213(89)90286-1} {\bibfield  {journal} {\bibinfo
  {journal} {Nucl. Phys. B}\ }\textbf {\bibinfo {volume} {327}},\ \bibinfo
  {pages} {49} (\bibinfo {year} {1989})},\ \bibinfo {note} {[Erratum:
  Nucl.Phys.B 335, 260--260 (1990)]}\BibitemShut {NoStop}%
\bibitem [{\citenamefont {Beenakker}\ \emph {et~al.}(1989)\citenamefont
  {Beenakker}, \citenamefont {Kuijf}, \citenamefont {van Neerven},\ and\
  \citenamefont {Smith}}]{Beenakker:1988bq}%
  \BibitemOpen
  \bibfield  {author} {\bibinfo {author} {\bibfnamefont {W.}~\bibnamefont
  {Beenakker}}, \bibinfo {author} {\bibfnamefont {H.}~\bibnamefont {Kuijf}},
  \bibinfo {author} {\bibfnamefont {W.~L.}\ \bibnamefont {van Neerven}}\ and\
  \bibinfo {author} {\bibfnamefont {J.}~\bibnamefont {Smith}},\ }\href
  {\doibase 10.1103/PhysRevD.40.54} {\bibfield  {journal} {\bibinfo  {journal}
  {Phys. Rev. D}\ }\textbf {\bibinfo {volume} {40}},\ \bibinfo {pages} {54}
  (\bibinfo {year} {1989})}\BibitemShut {NoStop}%
\bibitem [{\citenamefont {Mangano}\ \emph {et~al.}(1992)\citenamefont
  {Mangano}, \citenamefont {Nason},\ and\ \citenamefont
  {Ridolfi}}]{Mangano:1991jk}%
  \BibitemOpen
  \bibfield  {author} {\bibinfo {author} {\bibfnamefont {M.~L.}\ \bibnamefont
  {Mangano}}, \bibinfo {author} {\bibfnamefont {P.}~\bibnamefont {Nason}}\ and\
  \bibinfo {author} {\bibfnamefont {G.}~\bibnamefont {Ridolfi}},\ }\href
  {\doibase 10.1016/0550-3213(92)90435-E} {\bibfield  {journal} {\bibinfo
  {journal} {Nucl. Phys. B}\ }\textbf {\bibinfo {volume} {373}},\ \bibinfo
  {pages} {295} (\bibinfo {year} {1992})}\BibitemShut {NoStop}%
\bibitem [{\citenamefont {Garzelli}\ \emph {et~al.}(2021)\citenamefont
  {Garzelli}, \citenamefont {Kemmler}, \citenamefont {Moch},\ and\
  \citenamefont {Zenaiev}}]{Garzelli:2020fmd}%
  \BibitemOpen
  \bibfield  {author} {\bibinfo {author} {\bibfnamefont {M.~V.}\ \bibnamefont
  {Garzelli}}, \bibinfo {author} {\bibfnamefont {L.}~\bibnamefont {Kemmler}},
  \bibinfo {author} {\bibfnamefont {S.}~\bibnamefont {Moch}}\ and\ \bibinfo
  {author} {\bibfnamefont {O.}~\bibnamefont {Zenaiev}},\ }\href {\doibase
  10.1007/JHEP04(2021)043} {\bibfield  {journal} {\bibinfo  {journal} {JHEP}\
  }\textbf {\bibinfo {volume} {04}},\ \bibinfo {pages} {043} (\bibinfo {year}
  {2021})},\ \Eprint {http://arxiv.org/abs/2009.07763} {arXiv:2009.07763
  [hep-ph]}\BibitemShut {NoStop}%
\bibitem [{\citenamefont {Cacciari}\ and\ \citenamefont
  {Greco}(1994)}]{Cacciari:1993mq}%
  \BibitemOpen
  \bibfield  {author} {\bibinfo {author} {\bibfnamefont {M.}~\bibnamefont
  {Cacciari}}\ and\ \bibinfo {author} {\bibfnamefont {M.}~\bibnamefont
  {Greco}},\ }\href {\doibase 10.1016/0550-3213(94)90515-0} {\bibfield
  {journal} {\bibinfo  {journal} {Nucl. Phys. B}\ }\textbf {\bibinfo {volume}
  {421}},\ \bibinfo {pages} {530} (\bibinfo {year} {1994})},\ \Eprint
  {http://arxiv.org/abs/hep-ph/9311260} {arXiv:hep-ph/9311260}\BibitemShut
  {NoStop}%
\bibitem [{\citenamefont {Mele}\ and\ \citenamefont
  {Nason}(1991)}]{Mele:1990cw}%
  \BibitemOpen
  \bibfield  {author} {\bibinfo {author} {\bibfnamefont {B.}~\bibnamefont
  {Mele}}\ and\ \bibinfo {author} {\bibfnamefont {P.}~\bibnamefont {Nason}},\
  }\href {\doibase 10.1016/0550-3213(91)90597-Q} {\bibfield  {journal}
  {\bibinfo  {journal} {Nucl. Phys. B}\ }\textbf {\bibinfo {volume} {361}},\
  \bibinfo {pages} {626} (\bibinfo {year} {1991})},\ \bibinfo {note} {[Erratum:
  Nucl.Phys.B 921, 841--842 (2017)]}\BibitemShut {NoStop}%
\bibitem [{\citenamefont {Cacciari}\ and\ \citenamefont
  {Catani}(2001)}]{Cacciari:2001cw}%
  \BibitemOpen
  \bibfield  {author} {\bibinfo {author} {\bibfnamefont {M.}~\bibnamefont
  {Cacciari}}\ and\ \bibinfo {author} {\bibfnamefont {S.}~\bibnamefont
  {Catani}},\ }\href {\doibase 10.1016/S0550-3213(01)00469-2} {\bibfield
  {journal} {\bibinfo  {journal} {Nucl. Phys. B}\ }\textbf {\bibinfo {volume}
  {617}},\ \bibinfo {pages} {253} (\bibinfo {year} {2001})},\ \Eprint
  {http://arxiv.org/abs/hep-ph/0107138} {arXiv:hep-ph/0107138}\BibitemShut
  {NoStop}%
\bibitem [{\citenamefont {Kniehl}\ \emph
  {et~al.}(2005{\natexlab{a}})\citenamefont {Kniehl}, \citenamefont {Kramer},
  \citenamefont {Schienbein},\ and\ \citenamefont
  {Spiesberger}}]{Kniehl:2004fy}%
  \BibitemOpen
  \bibfield  {author} {\bibinfo {author} {\bibfnamefont {B.~A.}\ \bibnamefont
  {Kniehl}}, \bibinfo {author} {\bibfnamefont {G.}~\bibnamefont {Kramer}},
  \bibinfo {author} {\bibfnamefont {I.}~\bibnamefont {Schienbein}}\ and\
  \bibinfo {author} {\bibfnamefont {H.}~\bibnamefont {Spiesberger}},\ }\href
  {\doibase 10.1103/PhysRevD.71.014018} {\bibfield  {journal} {\bibinfo
  {journal} {Phys. Rev. D}\ }\textbf {\bibinfo {volume} {71}},\ \bibinfo
  {pages} {014018} (\bibinfo {year} {2005}{\natexlab{a}})},\ \Eprint
  {http://arxiv.org/abs/hep-ph/0410289} {arXiv:hep-ph/0410289}\BibitemShut
  {NoStop}%
\bibitem [{\citenamefont {Kniehl}\ \emph
  {et~al.}(2005{\natexlab{b}})\citenamefont {Kniehl}, \citenamefont {Kramer},
  \citenamefont {Schienbein},\ and\ \citenamefont
  {Spiesberger}}]{Kniehl:2005mk}%
  \BibitemOpen
  \bibfield  {author} {\bibinfo {author} {\bibfnamefont {B.~A.}\ \bibnamefont
  {Kniehl}}, \bibinfo {author} {\bibfnamefont {G.}~\bibnamefont {Kramer}},
  \bibinfo {author} {\bibfnamefont {I.}~\bibnamefont {Schienbein}}\ and\
  \bibinfo {author} {\bibfnamefont {H.}~\bibnamefont {Spiesberger}},\ }\href
  {\doibase 10.1140/epjc/s2005-02200-7} {\bibfield  {journal} {\bibinfo
  {journal} {Eur. Phys. J. C}\ }\textbf {\bibinfo {volume} {41}},\ \bibinfo
  {pages} {199} (\bibinfo {year} {2005}{\natexlab{b}})},\ \Eprint
  {http://arxiv.org/abs/hep-ph/0502194} {arXiv:hep-ph/0502194}\BibitemShut
  {NoStop}%
\bibitem [{\citenamefont {Kramer}\ and\ \citenamefont
  {Spiesberger}(2018)}]{Kramer:2018vde}%
  \BibitemOpen
  \bibfield  {author} {\bibinfo {author} {\bibfnamefont {G.}~\bibnamefont
  {Kramer}}\ and\ \bibinfo {author} {\bibfnamefont {H.}~\bibnamefont
  {Spiesberger}},\ }\href {\doibase 10.1103/PhysRevD.98.114010} {\bibfield
  {journal} {\bibinfo  {journal} {Phys. Rev. D}\ }\textbf {\bibinfo {volume}
  {98}},\ \bibinfo {pages} {114010} (\bibinfo {year} {2018})},\ \Eprint
  {http://arxiv.org/abs/1809.04297} {arXiv:1809.04297 [hep-ph]}\BibitemShut
  {NoStop}%
\bibitem [{\citenamefont {Benzke}\ \emph {et~al.}(2019)\citenamefont {Benzke},
  \citenamefont {Kniehl}, \citenamefont {Kramer}, \citenamefont {Schienbein},\
  and\ \citenamefont {Spiesberger}}]{Benzke:2019usl}%
  \BibitemOpen
  \bibfield  {author} {\bibinfo {author} {\bibfnamefont {M.}~\bibnamefont
  {Benzke}}, \bibinfo {author} {\bibfnamefont {B.~A.}\ \bibnamefont {Kniehl}},
  \bibinfo {author} {\bibfnamefont {G.}~\bibnamefont {Kramer}}, \bibinfo
  {author} {\bibfnamefont {I.}~\bibnamefont {Schienbein}}\ and\ \bibinfo
  {author} {\bibfnamefont {H.}~\bibnamefont {Spiesberger}},\ }\href {\doibase
  10.1140/epjc/s10052-019-7326-y} {\bibfield  {journal} {\bibinfo  {journal}
  {Eur. Phys. J. C}\ }\textbf {\bibinfo {volume} {79}},\ \bibinfo {pages} {814}
  (\bibinfo {year} {2019})},\ \Eprint {http://arxiv.org/abs/1907.12456}
  {arXiv:1907.12456 [hep-ph]}\BibitemShut {NoStop}%
\bibitem [{\citenamefont {Cacciari}\ \emph {et~al.}(1998)\citenamefont
  {Cacciari}, \citenamefont {Greco},\ and\ \citenamefont
  {Nason}}]{Cacciari:1998it}%
  \BibitemOpen
  \bibfield  {author} {\bibinfo {author} {\bibfnamefont {M.}~\bibnamefont
  {Cacciari}}, \bibinfo {author} {\bibfnamefont {M.}~\bibnamefont {Greco}}\
  and\ \bibinfo {author} {\bibfnamefont {P.}~\bibnamefont {Nason}},\ }\href
  {\doibase 10.1088/1126-6708/1998/05/007} {\bibfield  {journal} {\bibinfo
  {journal} {JHEP}\ }\textbf {\bibinfo {volume} {05}},\ \bibinfo {pages} {007}
  (\bibinfo {year} {1998})},\ \Eprint {http://arxiv.org/abs/hep-ph/9803400}
  {arXiv:hep-ph/9803400}\BibitemShut {NoStop}%
\bibitem [{\citenamefont {Cacciari}\ \emph {et~al.}(2001)\citenamefont
  {Cacciari}, \citenamefont {Frixione},\ and\ \citenamefont
  {Nason}}]{Cacciari:2001td}%
  \BibitemOpen
  \bibfield  {author} {\bibinfo {author} {\bibfnamefont {M.}~\bibnamefont
  {Cacciari}}, \bibinfo {author} {\bibfnamefont {S.}~\bibnamefont {Frixione}}\
  and\ \bibinfo {author} {\bibfnamefont {P.}~\bibnamefont {Nason}},\ }\href
  {\doibase 10.1088/1126-6708/2001/03/006} {\bibfield  {journal} {\bibinfo
  {journal} {JHEP}\ }\textbf {\bibinfo {volume} {03}},\ \bibinfo {pages} {006}
  (\bibinfo {year} {2001})},\ \Eprint {http://arxiv.org/abs/hep-ph/0102134}
  {arXiv:hep-ph/0102134}\BibitemShut {NoStop}%
\bibitem [{\citenamefont {Cacciari}\ and\ \citenamefont
  {Nason}(2002)}]{Cacciari:2002pa}%
  \BibitemOpen
  \bibfield  {author} {\bibinfo {author} {\bibfnamefont {M.}~\bibnamefont
  {Cacciari}}\ and\ \bibinfo {author} {\bibfnamefont {P.}~\bibnamefont
  {Nason}},\ }\href {\doibase 10.1103/PhysRevLett.89.122003} {\bibfield
  {journal} {\bibinfo  {journal} {Phys. Rev. Lett.}\ }\textbf {\bibinfo
  {volume} {89}},\ \bibinfo {pages} {122003} (\bibinfo {year} {2002})},\
  \Eprint {http://arxiv.org/abs/hep-ph/0204025}
  {arXiv:hep-ph/0204025}\BibitemShut {NoStop}%
\bibitem [{\citenamefont {Cacciari}\ \emph {et~al.}(2012)\citenamefont
  {Cacciari}, \citenamefont {Frixione}, \citenamefont {Houdeau}, \citenamefont
  {Mangano}, \citenamefont {Nason},\ and\ \citenamefont
  {Ridolfi}}]{Cacciari:2012ny}%
  \BibitemOpen
  \bibfield  {author} {\bibinfo {author} {\bibfnamefont {M.}~\bibnamefont
  {Cacciari}}, \bibinfo {author} {\bibfnamefont {S.}~\bibnamefont {Frixione}},
  \bibinfo {author} {\bibfnamefont {N.}~\bibnamefont {Houdeau}}, \bibinfo
  {author} {\bibfnamefont {M.~L.}\ \bibnamefont {Mangano}}, \bibinfo {author}
  {\bibfnamefont {P.}~\bibnamefont {Nason}}\ and\ \bibinfo {author}
  {\bibfnamefont {G.}~\bibnamefont {Ridolfi}},\ }\href {\doibase
  10.1007/JHEP10(2012)137} {\bibfield  {journal} {\bibinfo  {journal} {JHEP}\
  }\textbf {\bibinfo {volume} {10}},\ \bibinfo {pages} {137} (\bibinfo {year}
  {2012})},\ \Eprint {http://arxiv.org/abs/1205.6344} {arXiv:1205.6344
  [hep-ph]}\BibitemShut {NoStop}%
\bibitem [{\citenamefont {Cacciari}\ \emph {et~al.}(2006)\citenamefont
  {Cacciari}, \citenamefont {Nason},\ and\ \citenamefont
  {Oleari}}]{Cacciari:2005uk}%
  \BibitemOpen
  \bibfield  {author} {\bibinfo {author} {\bibfnamefont {M.}~\bibnamefont
  {Cacciari}}, \bibinfo {author} {\bibfnamefont {P.}~\bibnamefont {Nason}}\
  and\ \bibinfo {author} {\bibfnamefont {C.}~\bibnamefont {Oleari}},\ }\href
  {\doibase 10.1088/1126-6708/2006/04/006} {\bibfield  {journal} {\bibinfo
  {journal} {JHEP}\ }\textbf {\bibinfo {volume} {04}},\ \bibinfo {pages} {006}
  (\bibinfo {year} {2006})},\ \Eprint {http://arxiv.org/abs/hep-ph/0510032}
  {arXiv:hep-ph/0510032}\BibitemShut {NoStop}%
\bibitem [{\citenamefont {Kartvelishvili}\ \emph {et~al.}(1978)\citenamefont
  {Kartvelishvili}, \citenamefont {Likhoded},\ and\ \citenamefont
  {Petrov}}]{Kartvelishvili:1977pi}%
  \BibitemOpen
  \bibfield  {author} {\bibinfo {author} {\bibfnamefont {V.~G.}\ \bibnamefont
  {Kartvelishvili}}, \bibinfo {author} {\bibfnamefont {A.~K.}\ \bibnamefont
  {Likhoded}}\ and\ \bibinfo {author} {\bibfnamefont {V.~A.}\ \bibnamefont
  {Petrov}},\ }\href {\doibase 10.1016/0370-2693(78)90653-6} {\bibfield
  {journal} {\bibinfo  {journal} {Phys. Lett. B}\ }\textbf {\bibinfo {volume}
  {78}},\ \bibinfo {pages} {615} (\bibinfo {year} {1978})}\BibitemShut
  {NoStop}%
\bibitem [{\citenamefont {Peterson}\ \emph {et~al.}(1983)\citenamefont
  {Peterson}, \citenamefont {Schlatter}, \citenamefont {Schmitt},\ and\
  \citenamefont {Zerwas}}]{Peterson:1982ak}%
  \BibitemOpen
  \bibfield  {author} {\bibinfo {author} {\bibfnamefont {C.}~\bibnamefont
  {Peterson}}, \bibinfo {author} {\bibfnamefont {D.}~\bibnamefont {Schlatter}},
  \bibinfo {author} {\bibfnamefont {I.}~\bibnamefont {Schmitt}}\ and\ \bibinfo
  {author} {\bibfnamefont {P.~M.}\ \bibnamefont {Zerwas}},\ }\href {\doibase
  10.1103/PhysRevD.27.105} {\bibfield  {journal} {\bibinfo  {journal} {Phys.
  Rev. D}\ }\textbf {\bibinfo {volume} {27}},\ \bibinfo {pages} {105} (\bibinfo
  {year} {1983})}\BibitemShut {NoStop}%
\bibitem [{\citenamefont {Frixione}\ \emph
  {et~al.}(2007{\natexlab{a}})\citenamefont {Frixione}, \citenamefont {Nason},\
  and\ \citenamefont {Ridolfi}}]{Frixione:2007nw}%
  \BibitemOpen
  \bibfield  {author} {\bibinfo {author} {\bibfnamefont {S.}~\bibnamefont
  {Frixione}}, \bibinfo {author} {\bibfnamefont {P.}~\bibnamefont {Nason}}\
  and\ \bibinfo {author} {\bibfnamefont {G.}~\bibnamefont {Ridolfi}},\ }\href
  {\doibase 10.1088/1126-6708/2007/09/126} {\bibfield  {journal} {\bibinfo
  {journal} {JHEP}\ }\textbf {\bibinfo {volume} {09}},\ \bibinfo {pages} {126}
  (\bibinfo {year} {2007}{\natexlab{a}})},\ \Eprint
  {http://arxiv.org/abs/0707.3088} {arXiv:0707.3088 [hep-ph]}\BibitemShut
  {NoStop}%
\bibitem [{\citenamefont {Buonocore}\ \emph {et~al.}(2018)\citenamefont
  {Buonocore}, \citenamefont {Nason},\ and\ \citenamefont
  {Tramontano}}]{Buonocore:2017lry}%
  \BibitemOpen
  \bibfield  {author} {\bibinfo {author} {\bibfnamefont {L.}~\bibnamefont
  {Buonocore}}, \bibinfo {author} {\bibfnamefont {P.}~\bibnamefont {Nason}}\
  and\ \bibinfo {author} {\bibfnamefont {F.}~\bibnamefont {Tramontano}},\
  }\href {\doibase 10.1140/epjc/s10052-018-5638-y} {\bibfield  {journal}
  {\bibinfo  {journal} {Eur. Phys. J. C}\ }\textbf {\bibinfo {volume} {78}},\
  \bibinfo {pages} {151} (\bibinfo {year} {2018})},\ \Eprint
  {http://arxiv.org/abs/1711.06281} {arXiv:1711.06281 [hep-ph]}\BibitemShut
  {NoStop}%
\bibitem [{\citenamefont {Alwall}\ \emph {et~al.}(2014)\citenamefont {Alwall},
  \citenamefont {Frederix}, \citenamefont {Frixione}, \citenamefont {Hirschi},
  \citenamefont {Maltoni}, \citenamefont {Mattelaer}, \citenamefont {Shao},
  \citenamefont {Stelzer}, \citenamefont {Torrielli},\ and\ \citenamefont
  {Zaro}}]{Alwall:2014hca}%
  \BibitemOpen
  \bibfield  {author} {\bibinfo {author} {\bibfnamefont {J.}~\bibnamefont
  {Alwall}}, \bibinfo {author} {\bibfnamefont {R.}~\bibnamefont {Frederix}},
  \bibinfo {author} {\bibfnamefont {S.}~\bibnamefont {Frixione}}, \bibinfo
  {author} {\bibfnamefont {V.}~\bibnamefont {Hirschi}}, \bibinfo {author}
  {\bibfnamefont {F.}~\bibnamefont {Maltoni}}, \bibinfo {author} {\bibfnamefont
  {O.}~\bibnamefont {Mattelaer}}, \bibinfo {author} {\bibfnamefont {H.~S.}\
  \bibnamefont {Shao}}, \bibinfo {author} {\bibfnamefont {T.}~\bibnamefont
  {Stelzer}}, \bibinfo {author} {\bibfnamefont {P.}~\bibnamefont {Torrielli}}\
  and\ \bibinfo {author} {\bibfnamefont {M.}~\bibnamefont {Zaro}},\ }\href
  {\doibase 10.1007/JHEP07(2014)079} {\bibfield  {journal} {\bibinfo  {journal}
  {JHEP}\ }\textbf {\bibinfo {volume} {07}},\ \bibinfo {pages} {079} (\bibinfo
  {year} {2014})},\ \Eprint {http://arxiv.org/abs/1405.0301} {arXiv:1405.0301
  [hep-ph]}\BibitemShut {NoStop}%
\bibitem [{\citenamefont {Mangano}\ \emph {et~al.}(2016)\citenamefont {Mangano}
  \emph {et~al.}}]{Mangano:2016jyj}%
  \BibitemOpen
  \bibfield  {author} {\bibinfo {author} {\bibfnamefont {M.~L.}\ \bibnamefont
  {Mangano}} \emph {et~al.},\ }\href {\doibase 10.23731/CYRM-2017-003.1} {\
  (\bibinfo {year} {2016}),\ 10.23731/CYRM-2017-003.1},\ \Eprint
  {http://arxiv.org/abs/1607.01831} {arXiv:1607.01831 [hep-ph]}\BibitemShut
  {NoStop}%
\bibitem [{\citenamefont {d'Enterria}\ and\ \citenamefont
  {Snigirev}(2017)}]{dEnterria:2016ids}%
  \BibitemOpen
  \bibfield  {author} {\bibinfo {author} {\bibfnamefont {D.}~\bibnamefont
  {d'Enterria}}\ and\ \bibinfo {author} {\bibfnamefont {A.~M.}\ \bibnamefont
  {Snigirev}},\ }\href {\doibase 10.1103/PhysRevLett.118.122001} {\bibfield
  {journal} {\bibinfo  {journal} {Phys. Rev. Lett.}\ }\textbf {\bibinfo
  {volume} {118}},\ \bibinfo {pages} {122001} (\bibinfo {year} {2017})},\
  \Eprint {http://arxiv.org/abs/1612.05582} {arXiv:1612.05582
  [hep-ph]}\BibitemShut {NoStop}%
\bibitem [{\citenamefont {Langenfeld}\ \emph {et~al.}(2009)\citenamefont
  {Langenfeld}, \citenamefont {Moch},\ and\ \citenamefont
  {Uwer}}]{Langenfeld:2009wd}%
  \BibitemOpen
  \bibfield  {author} {\bibinfo {author} {\bibfnamefont {U.}~\bibnamefont
  {Langenfeld}}, \bibinfo {author} {\bibfnamefont {S.}~\bibnamefont {Moch}}\
  and\ \bibinfo {author} {\bibfnamefont {P.}~\bibnamefont {Uwer}},\ }\href
  {\doibase 10.1103/PhysRevD.80.054009} {\bibfield  {journal} {\bibinfo
  {journal} {Phys. Rev. D}\ }\textbf {\bibinfo {volume} {80}},\ \bibinfo
  {pages} {054009} (\bibinfo {year} {2009})},\ \Eprint
  {http://arxiv.org/abs/0906.5273} {arXiv:0906.5273 [hep-ph]}\BibitemShut
  {NoStop}%
\bibitem [{\citenamefont {Aliev}\ \emph {et~al.}(2011)\citenamefont {Aliev},
  \citenamefont {Lacker}, \citenamefont {Langenfeld}, \citenamefont {Moch},
  \citenamefont {Uwer},\ and\ \citenamefont {Wiedermann}}]{Aliev:2010zk}%
  \BibitemOpen
  \bibfield  {author} {\bibinfo {author} {\bibfnamefont {M.}~\bibnamefont
  {Aliev}}, \bibinfo {author} {\bibfnamefont {H.}~\bibnamefont {Lacker}},
  \bibinfo {author} {\bibfnamefont {U.}~\bibnamefont {Langenfeld}}, \bibinfo
  {author} {\bibfnamefont {S.}~\bibnamefont {Moch}}, \bibinfo {author}
  {\bibfnamefont {P.}~\bibnamefont {Uwer}}\ and\ \bibinfo {author}
  {\bibfnamefont {M.}~\bibnamefont {Wiedermann}},\ }\href {\doibase
  10.1016/j.cpc.2010.12.040} {\bibfield  {journal} {\bibinfo  {journal}
  {Comput. Phys. Commun.}\ }\textbf {\bibinfo {volume} {182}},\ \bibinfo
  {pages} {1034} (\bibinfo {year} {2011})},\ \Eprint
  {http://arxiv.org/abs/1007.1327} {arXiv:1007.1327 [hep-ph]}\BibitemShut
  {NoStop}%
\bibitem [{\citenamefont {Catani}\ \emph {et~al.}(2021)\citenamefont {Catani},
  \citenamefont {Devoto}, \citenamefont {Grazzini}, \citenamefont {Kallweit},\
  and\ \citenamefont {Mazzitelli}}]{Catani:2020kkl}%
  \BibitemOpen
  \bibfield  {author} {\bibinfo {author} {\bibfnamefont {S.}~\bibnamefont
  {Catani}}, \bibinfo {author} {\bibfnamefont {S.}~\bibnamefont {Devoto}},
  \bibinfo {author} {\bibfnamefont {M.}~\bibnamefont {Grazzini}}, \bibinfo
  {author} {\bibfnamefont {S.}~\bibnamefont {Kallweit}}\ and\ \bibinfo {author}
  {\bibfnamefont {J.}~\bibnamefont {Mazzitelli}},\ }\href {\doibase
  10.1007/JHEP03(2021)029} {\bibfield  {journal} {\bibinfo  {journal} {JHEP}\
  }\textbf {\bibinfo {volume} {03}},\ \bibinfo {pages} {029} (\bibinfo {year}
  {2021})},\ \Eprint {http://arxiv.org/abs/2010.11906} {arXiv:2010.11906
  [hep-ph]}\BibitemShut {NoStop}%
\bibitem [{\citenamefont {Grazzini}\ \emph {et~al.}(2018)\citenamefont
  {Grazzini}, \citenamefont {Kallweit},\ and\ \citenamefont
  {Wiesemann}}]{Grazzini:2017mhc}%
  \BibitemOpen
  \bibfield  {author} {\bibinfo {author} {\bibfnamefont {M.}~\bibnamefont
  {Grazzini}}, \bibinfo {author} {\bibfnamefont {S.}~\bibnamefont {Kallweit}}\
  and\ \bibinfo {author} {\bibfnamefont {M.}~\bibnamefont {Wiesemann}},\ }\href
  {\doibase 10.1140/epjc/s10052-018-5771-7} {\bibfield  {journal} {\bibinfo
  {journal} {Eur. Phys. J.}\ }\textbf {\bibinfo {volume} {C78}},\ \bibinfo
  {pages} {537} (\bibinfo {year} {2018})},\ \Eprint
  {http://arxiv.org/abs/1711.06631} {arXiv:1711.06631 [hep-ph]}\BibitemShut
  {NoStop}%
%%CITATION = ARXIV:1711.06631;%%
\bibitem [{\citenamefont {Czakon}\ \emph {et~al.}(2021)\citenamefont {Czakon},
  \citenamefont {Generet}, \citenamefont {Mitov},\ and\ \citenamefont
  {Poncelet}}]{Czakon:2021ohs}%
  \BibitemOpen
  \bibfield  {author} {\bibinfo {author} {\bibfnamefont {M.~L.}\ \bibnamefont
  {Czakon}}, \bibinfo {author} {\bibfnamefont {T.}~\bibnamefont {Generet}},
  \bibinfo {author} {\bibfnamefont {A.}~\bibnamefont {Mitov}}\ and\ \bibinfo
  {author} {\bibfnamefont {R.}~\bibnamefont {Poncelet}},\ }\href {\doibase
  10.1007/JHEP10(2021)216} {\bibfield  {journal} {\bibinfo  {journal} {JHEP}\
  }\textbf {\bibinfo {volume} {10}},\ \bibinfo {pages} {216} (\bibinfo {year}
  {2021})},\ \Eprint {http://arxiv.org/abs/2102.08267} {arXiv:2102.08267
  [hep-ph]}\BibitemShut {NoStop}%
\bibitem [{\citenamefont {Mazzitelli}\ \emph {et~al.}(2021)\citenamefont
  {Mazzitelli}, \citenamefont {Monni}, \citenamefont {Nason}, \citenamefont
  {Re}, \citenamefont {Wiesemann},\ and\ \citenamefont
  {Zanderighi}}]{Mazzitelli:2020jio}%
  \BibitemOpen
  \bibfield  {author} {\bibinfo {author} {\bibfnamefont {J.}~\bibnamefont
  {Mazzitelli}}, \bibinfo {author} {\bibfnamefont {P.~F.}\ \bibnamefont
  {Monni}}, \bibinfo {author} {\bibfnamefont {P.}~\bibnamefont {Nason}},
  \bibinfo {author} {\bibfnamefont {E.}~\bibnamefont {Re}}, \bibinfo {author}
  {\bibfnamefont {M.}~\bibnamefont {Wiesemann}}\ and\ \bibinfo {author}
  {\bibfnamefont {G.}~\bibnamefont {Zanderighi}},\ }\href {\doibase
  10.1103/PhysRevLett.127.062001} {\bibfield  {journal} {\bibinfo  {journal}
  {Phys. Rev. Lett.}\ }\textbf {\bibinfo {volume} {127}},\ \bibinfo {pages}
  {062001} (\bibinfo {year} {2021})},\ \Eprint
  {http://arxiv.org/abs/2012.14267} {arXiv:2012.14267 [hep-ph]}\BibitemShut
  {NoStop}%
\bibitem [{\citenamefont {Mazzitelli}\ \emph {et~al.}(2022)\citenamefont
  {Mazzitelli}, \citenamefont {Monni}, \citenamefont {Nason}, \citenamefont
  {Re}, \citenamefont {Wiesemann},\ and\ \citenamefont
  {Zanderighi}}]{Mazzitelli:2021mmm}%
  \BibitemOpen
  \bibfield  {author} {\bibinfo {author} {\bibfnamefont {J.}~\bibnamefont
  {Mazzitelli}}, \bibinfo {author} {\bibfnamefont {P.~F.}\ \bibnamefont
  {Monni}}, \bibinfo {author} {\bibfnamefont {P.}~\bibnamefont {Nason}},
  \bibinfo {author} {\bibfnamefont {E.}~\bibnamefont {Re}}, \bibinfo {author}
  {\bibfnamefont {M.}~\bibnamefont {Wiesemann}}\ and\ \bibinfo {author}
  {\bibfnamefont {G.}~\bibnamefont {Zanderighi}},\ }\href {\doibase
  10.1007/JHEP04(2022)079} {\bibfield  {journal} {\bibinfo  {journal} {JHEP}\
  }\textbf {\bibinfo {volume} {04}},\ \bibinfo {pages} {079} (\bibinfo {year}
  {2022})},\ \Eprint {http://arxiv.org/abs/2112.12135} {arXiv:2112.12135
  [hep-ph]}\BibitemShut {NoStop}%
\bibitem [{\citenamefont {Monni}\ \emph
  {et~al.}(2020{\natexlab{a}})\citenamefont {Monni}, \citenamefont {Nason},
  \citenamefont {Re}, \citenamefont {Wiesemann},\ and\ \citenamefont
  {Zanderighi}}]{Monni:2019whf}%
  \BibitemOpen
  \bibfield  {author} {\bibinfo {author} {\bibfnamefont {P.~F.}\ \bibnamefont
  {Monni}}, \bibinfo {author} {\bibfnamefont {P.}~\bibnamefont {Nason}},
  \bibinfo {author} {\bibfnamefont {E.}~\bibnamefont {Re}}, \bibinfo {author}
  {\bibfnamefont {M.}~\bibnamefont {Wiesemann}}\ and\ \bibinfo {author}
  {\bibfnamefont {G.}~\bibnamefont {Zanderighi}},\ }\href {\doibase
  10.1007/JHEP05(2020)143} {\bibfield  {journal} {\bibinfo  {journal} {JHEP}\
  }\textbf {\bibinfo {volume} {05}},\ \bibinfo {pages} {143} (\bibinfo {year}
  {2020}{\natexlab{a}})},\ \Eprint {http://arxiv.org/abs/1908.06987}
  {arXiv:1908.06987 [hep-ph]}\BibitemShut {NoStop}%
\bibitem [{\citenamefont {Monni}\ \emph
  {et~al.}(2020{\natexlab{b}})\citenamefont {Monni}, \citenamefont {Re},\ and\
  \citenamefont {Wiesemann}}]{Monni:2020nks}%
  \BibitemOpen
  \bibfield  {author} {\bibinfo {author} {\bibfnamefont {P.~F.}\ \bibnamefont
  {Monni}}, \bibinfo {author} {\bibfnamefont {E.}~\bibnamefont {Re}}\ and\
  \bibinfo {author} {\bibfnamefont {M.}~\bibnamefont {Wiesemann}},\ }\href
  {\doibase 10.1140/epjc/s10052-020-08658-5} {\bibfield  {journal} {\bibinfo
  {journal} {Eur. Phys. J. C}\ }\textbf {\bibinfo {volume} {80}},\ \bibinfo
  {pages} {1075} (\bibinfo {year} {2020}{\natexlab{b}})},\ \Eprint
  {http://arxiv.org/abs/2006.04133} {arXiv:2006.04133 [hep-ph]}\BibitemShut
  {NoStop}%
\bibitem [{\citenamefont {Lombardi}\ \emph
  {et~al.}(2021{\natexlab{a}})\citenamefont {Lombardi}, \citenamefont
  {Wiesemann},\ and\ \citenamefont {Zanderighi}}]{Lombardi:2020wju}%
  \BibitemOpen
  \bibfield  {author} {\bibinfo {author} {\bibfnamefont {D.}~\bibnamefont
  {Lombardi}}, \bibinfo {author} {\bibfnamefont {M.}~\bibnamefont {Wiesemann}}\
  and\ \bibinfo {author} {\bibfnamefont {G.}~\bibnamefont {Zanderighi}},\
  }\href {\doibase 10.1007/JHEP06(2021)095} {\bibfield  {journal} {\bibinfo
  {journal} {JHEP}\ }\textbf {\bibinfo {volume} {06}},\ \bibinfo {pages} {095}
  (\bibinfo {year} {2021}{\natexlab{a}})},\ \Eprint
  {http://arxiv.org/abs/2010.10478} {arXiv:2010.10478 [hep-ph]}\BibitemShut
  {NoStop}%
\bibitem [{\citenamefont {Lombardi}\ \emph
  {et~al.}(2021{\natexlab{b}})\citenamefont {Lombardi}, \citenamefont
  {Wiesemann},\ and\ \citenamefont {Zanderighi}}]{Lombardi:2021rvg}%
  \BibitemOpen
  \bibfield  {author} {\bibinfo {author} {\bibfnamefont {D.}~\bibnamefont
  {Lombardi}}, \bibinfo {author} {\bibfnamefont {M.}~\bibnamefont {Wiesemann}}\
  and\ \bibinfo {author} {\bibfnamefont {G.}~\bibnamefont {Zanderighi}},\
  }\href {\doibase 10.1007/JHEP11(2021)230} {\bibfield  {journal} {\bibinfo
  {journal} {JHEP}\ }\textbf {\bibinfo {volume} {11}},\ \bibinfo {pages} {230}
  (\bibinfo {year} {2021}{\natexlab{b}})},\ \Eprint
  {http://arxiv.org/abs/2103.12077} {arXiv:2103.12077 [hep-ph]}\BibitemShut
  {NoStop}%
\bibitem [{\citenamefont {Buonocore}\ \emph {et~al.}(2022)\citenamefont
  {Buonocore}, \citenamefont {Koole}, \citenamefont {Lombardi}, \citenamefont
  {Rottoli}, \citenamefont {Wiesemann},\ and\ \citenamefont
  {Zanderighi}}]{Buonocore:2021fnj}%
  \BibitemOpen
  \bibfield  {author} {\bibinfo {author} {\bibfnamefont {L.}~\bibnamefont
  {Buonocore}}, \bibinfo {author} {\bibfnamefont {G.}~\bibnamefont {Koole}},
  \bibinfo {author} {\bibfnamefont {D.}~\bibnamefont {Lombardi}}, \bibinfo
  {author} {\bibfnamefont {L.}~\bibnamefont {Rottoli}}, \bibinfo {author}
  {\bibfnamefont {M.}~\bibnamefont {Wiesemann}}\ and\ \bibinfo {author}
  {\bibfnamefont {G.}~\bibnamefont {Zanderighi}},\ }\href {\doibase
  10.1007/JHEP01(2022)072} {\bibfield  {journal} {\bibinfo  {journal} {JHEP}\
  }\textbf {\bibinfo {volume} {01}},\ \bibinfo {pages} {072} (\bibinfo {year}
  {2022})},\ \Eprint {http://arxiv.org/abs/2108.05337} {arXiv:2108.05337
  [hep-ph]}\BibitemShut {NoStop}%
\bibitem [{\citenamefont {Lombardi}\ \emph {et~al.}(2022)\citenamefont
  {Lombardi}, \citenamefont {Wiesemann},\ and\ \citenamefont
  {Zanderighi}}]{Lombardi:2021wug}%
  \BibitemOpen
  \bibfield  {author} {\bibinfo {author} {\bibfnamefont {D.}~\bibnamefont
  {Lombardi}}, \bibinfo {author} {\bibfnamefont {M.}~\bibnamefont {Wiesemann}}\
  and\ \bibinfo {author} {\bibfnamefont {G.}~\bibnamefont {Zanderighi}},\
  }\href {\doibase 10.1016/j.physletb.2021.136846} {\bibfield  {journal}
  {\bibinfo  {journal} {Phys. Lett. B}\ }\textbf {\bibinfo {volume} {824}},\
  \bibinfo {pages} {136846} (\bibinfo {year} {2022})},\ \Eprint
  {http://arxiv.org/abs/2108.11315} {arXiv:2108.11315 [hep-ph]}\BibitemShut
  {NoStop}%
\bibitem [{\citenamefont {Zanoli}\ \emph {et~al.}(2022)\citenamefont {Zanoli},
  \citenamefont {Chiesa}, \citenamefont {Re}, \citenamefont {Wiesemann},\ and\
  \citenamefont {Zanderighi}}]{Zanoli:2021iyp}%
  \BibitemOpen
  \bibfield  {author} {\bibinfo {author} {\bibfnamefont {S.}~\bibnamefont
  {Zanoli}}, \bibinfo {author} {\bibfnamefont {M.}~\bibnamefont {Chiesa}},
  \bibinfo {author} {\bibfnamefont {E.}~\bibnamefont {Re}}, \bibinfo {author}
  {\bibfnamefont {M.}~\bibnamefont {Wiesemann}}\ and\ \bibinfo {author}
  {\bibfnamefont {G.}~\bibnamefont {Zanderighi}},\ }\href {\doibase
  10.1007/JHEP07(2022)008} {\bibfield  {journal} {\bibinfo  {journal} {JHEP}\
  }\textbf {\bibinfo {volume} {07}},\ \bibinfo {pages} {008} (\bibinfo {year}
  {2022})},\ \Eprint {http://arxiv.org/abs/2112.04168} {arXiv:2112.04168
  [hep-ph]}\BibitemShut {NoStop}%
\bibitem [{\citenamefont {Gavardi}\ \emph {et~al.}(2022)\citenamefont
  {Gavardi}, \citenamefont {Oleari},\ and\ \citenamefont
  {Re}}]{Gavardi:2022ixt}%
  \BibitemOpen
  \bibfield  {author} {\bibinfo {author} {\bibfnamefont {A.}~\bibnamefont
  {Gavardi}}, \bibinfo {author} {\bibfnamefont {C.}~\bibnamefont {Oleari}}\
  and\ \bibinfo {author} {\bibfnamefont {E.}~\bibnamefont {Re}},\ }\href
  {\doibase 10.1007/JHEP09(2022)061} {\bibfield  {journal} {\bibinfo  {journal}
  {JHEP}\ }\textbf {\bibinfo {volume} {09}},\ \bibinfo {pages} {061} (\bibinfo
  {year} {2022})},\ \Eprint {http://arxiv.org/abs/2204.12602} {arXiv:2204.12602
  [hep-ph]}\BibitemShut {NoStop}%
\bibitem [{\citenamefont {Haisch}\ \emph {et~al.}(2022)\citenamefont {Haisch},
  \citenamefont {Scott}, \citenamefont {Wiesemann}, \citenamefont
  {Zanderighi},\ and\ \citenamefont {Zanoli}}]{Haisch:2022nwz}%
  \BibitemOpen
  \bibfield  {author} {\bibinfo {author} {\bibfnamefont {U.}~\bibnamefont
  {Haisch}}, \bibinfo {author} {\bibfnamefont {D.~J.}\ \bibnamefont {Scott}},
  \bibinfo {author} {\bibfnamefont {M.}~\bibnamefont {Wiesemann}}, \bibinfo
  {author} {\bibfnamefont {G.}~\bibnamefont {Zanderighi}}\ and\ \bibinfo
  {author} {\bibfnamefont {S.}~\bibnamefont {Zanoli}},\ }\href {\doibase
  10.1007/JHEP07(2022)054} {\bibfield  {journal} {\bibinfo  {journal} {JHEP}\
  }\textbf {\bibinfo {volume} {07}},\ \bibinfo {pages} {054} (\bibinfo {year}
  {2022})},\ \Eprint {http://arxiv.org/abs/2204.00663} {arXiv:2204.00663
  [hep-ph]}\BibitemShut {NoStop}%
\bibitem [{\citenamefont {Lindert}\ \emph {et~al.}(2022)\citenamefont
  {Lindert}, \citenamefont {Lombardi}, \citenamefont {Wiesemann}, \citenamefont
  {Zanderighi},\ and\ \citenamefont {Zanoli}}]{Lindert:2022qdd}%
  \BibitemOpen
  \bibfield  {author} {\bibinfo {author} {\bibfnamefont {J.~M.}\ \bibnamefont
  {Lindert}}, \bibinfo {author} {\bibfnamefont {D.}~\bibnamefont {Lombardi}},
  \bibinfo {author} {\bibfnamefont {M.}~\bibnamefont {Wiesemann}}, \bibinfo
  {author} {\bibfnamefont {G.}~\bibnamefont {Zanderighi}}\ and\ \bibinfo
  {author} {\bibfnamefont {S.}~\bibnamefont {Zanoli}},\ }\href {\doibase
  10.1007/JHEP11(2022)036} {\bibfield  {journal} {\bibinfo  {journal} {JHEP}\
  }\textbf {\bibinfo {volume} {11}},\ \bibinfo {pages} {036} (\bibinfo {year}
  {2022})},\ \Eprint {http://arxiv.org/abs/2208.12660} {arXiv:2208.12660
  [hep-ph]}\BibitemShut {NoStop}%
\bibitem [{\citenamefont {Zhu}\ \emph {et~al.}(2013)\citenamefont {Zhu},
  \citenamefont {Li}, \citenamefont {Li}, \citenamefont {Shao},\ and\
  \citenamefont {Yang}}]{Zhu:2012ts}%
  \BibitemOpen
  \bibfield  {author} {\bibinfo {author} {\bibfnamefont {H.~X.}\ \bibnamefont
  {Zhu}}, \bibinfo {author} {\bibfnamefont {C.~S.}\ \bibnamefont {Li}},
  \bibinfo {author} {\bibfnamefont {H.~T.}\ \bibnamefont {Li}}, \bibinfo
  {author} {\bibfnamefont {D.~Y.}\ \bibnamefont {Shao}}\ and\ \bibinfo {author}
  {\bibfnamefont {L.~L.}\ \bibnamefont {Yang}},\ }\href {\doibase
  10.1103/PhysRevLett.110.082001} {\bibfield  {journal} {\bibinfo  {journal}
  {Phys. Rev. Lett.}\ }\textbf {\bibinfo {volume} {110}},\ \bibinfo {pages}
  {082001} (\bibinfo {year} {2013})},\ \Eprint {http://arxiv.org/abs/1208.5774}
  {arXiv:1208.5774 [hep-ph]}\BibitemShut {NoStop}%
\bibitem [{\citenamefont {Li}\ \emph {et~al.}(2013)\citenamefont {Li},
  \citenamefont {Li}, \citenamefont {Shao}, \citenamefont {Yang},\ and\
  \citenamefont {Zhu}}]{Li:2013mia}%
  \BibitemOpen
  \bibfield  {author} {\bibinfo {author} {\bibfnamefont {H.~T.}\ \bibnamefont
  {Li}}, \bibinfo {author} {\bibfnamefont {C.~S.}\ \bibnamefont {Li}}, \bibinfo
  {author} {\bibfnamefont {D.~Y.}\ \bibnamefont {Shao}}, \bibinfo {author}
  {\bibfnamefont {L.~L.}\ \bibnamefont {Yang}}\ and\ \bibinfo {author}
  {\bibfnamefont {H.~X.}\ \bibnamefont {Zhu}},\ }\href {\doibase
  10.1103/PhysRevD.88.074004} {\bibfield  {journal} {\bibinfo  {journal} {Phys.
  Rev. D}\ }\textbf {\bibinfo {volume} {88}},\ \bibinfo {pages} {074004}
  (\bibinfo {year} {2013})},\ \Eprint {http://arxiv.org/abs/1307.2464}
  {arXiv:1307.2464 [hep-ph]}\BibitemShut {NoStop}%
\bibitem [{\citenamefont {Catani}\ \emph {et~al.}(2014)\citenamefont {Catani},
  \citenamefont {Grazzini},\ and\ \citenamefont {Torre}}]{Catani:2014qha}%
  \BibitemOpen
  \bibfield  {author} {\bibinfo {author} {\bibfnamefont {S.}~\bibnamefont
  {Catani}}, \bibinfo {author} {\bibfnamefont {M.}~\bibnamefont {Grazzini}}\
  and\ \bibinfo {author} {\bibfnamefont {A.}~\bibnamefont {Torre}},\ }\href
  {\doibase 10.1016/j.nuclphysb.2014.11.019} {\bibfield  {journal} {\bibinfo
  {journal} {Nucl. Phys.}\ }\textbf {\bibinfo {volume} {B890}},\ \bibinfo
  {pages} {518} (\bibinfo {year} {2014})},\ \Eprint
  {http://arxiv.org/abs/1408.4564} {arXiv:1408.4564 [hep-ph]}\BibitemShut
  {NoStop}%
%%CITATION = ARXIV:1408.4564;%%
\bibitem [{\citenamefont {Catani}\ \emph {et~al.}(2018)\citenamefont {Catani},
  \citenamefont {Grazzini},\ and\ \citenamefont {Sargsyan}}]{Catani:2018mei}%
  \BibitemOpen
  \bibfield  {author} {\bibinfo {author} {\bibfnamefont {S.}~\bibnamefont
  {Catani}}, \bibinfo {author} {\bibfnamefont {M.}~\bibnamefont {Grazzini}}\
  and\ \bibinfo {author} {\bibfnamefont {H.}~\bibnamefont {Sargsyan}},\ }\href
  {\doibase 10.1007/JHEP11(2018)061} {\bibfield  {journal} {\bibinfo  {journal}
  {JHEP}\ }\textbf {\bibinfo {volume} {11}},\ \bibinfo {pages} {061} (\bibinfo
  {year} {2018})},\ \Eprint {http://arxiv.org/abs/1806.01601} {arXiv:1806.01601
  [hep-ph]}\BibitemShut {NoStop}%
\bibitem [{\citenamefont {Catani}\ \emph {et~al.}(2023)\citenamefont {Catani},
  \citenamefont {Devoto}, \citenamefont {Grazzini},\ and\ \citenamefont
  {Mazzitelli}}]{Catani:2023tby}%
  \BibitemOpen
  \bibfield  {author} {\bibinfo {author} {\bibfnamefont {S.}~\bibnamefont
  {Catani}}, \bibinfo {author} {\bibfnamefont {S.}~\bibnamefont {Devoto}},
  \bibinfo {author} {\bibfnamefont {M.}~\bibnamefont {Grazzini}}\ and\ \bibinfo
  {author} {\bibfnamefont {J.}~\bibnamefont {Mazzitelli}},\ }\href@noop {} {\
  (\bibinfo {year} {2023})},\ \Eprint {http://arxiv.org/abs/2301.11786}
  {arXiv:2301.11786 [hep-ph]}\BibitemShut {NoStop}%
\bibitem [{\citenamefont {Nason}(2004)}]{Nason:2004rx}%
  \BibitemOpen
  \bibfield  {author} {\bibinfo {author} {\bibfnamefont {P.}~\bibnamefont
  {Nason}},\ }\href {\doibase 10.1088/1126-6708/2004/11/040} {\bibfield
  {journal} {\bibinfo  {journal} {JHEP}\ }\textbf {\bibinfo {volume} {11}},\
  \bibinfo {pages} {040} (\bibinfo {year} {2004})},\ \Eprint
  {http://arxiv.org/abs/hep-ph/0409146} {arXiv:hep-ph/0409146
  [hep-ph]}\BibitemShut {NoStop}%
%%CITATION = HEP-PH/0409146;%%
\bibitem [{\citenamefont {Nason}\ and\ \citenamefont
  {Ridolfi}(2006)}]{Nason:2006hfa}%
  \BibitemOpen
  \bibfield  {author} {\bibinfo {author} {\bibfnamefont {P.}~\bibnamefont
  {Nason}}\ and\ \bibinfo {author} {\bibfnamefont {G.}~\bibnamefont
  {Ridolfi}},\ }\href {\doibase 10.1088/1126-6708/2006/08/077} {\bibfield
  {journal} {\bibinfo  {journal} {JHEP}\ }\textbf {\bibinfo {volume} {08}},\
  \bibinfo {pages} {077} (\bibinfo {year} {2006})},\ \Eprint
  {http://arxiv.org/abs/hep-ph/0606275} {arXiv:hep-ph/0606275
  [hep-ph]}\BibitemShut {NoStop}%
%%CITATION = HEP-PH/0606275;%%
\bibitem [{\citenamefont {Frixione}\ \emph
  {et~al.}(2007{\natexlab{b}})\citenamefont {Frixione}, \citenamefont {Nason},\
  and\ \citenamefont {Oleari}}]{Frixione:2007vw}%
  \BibitemOpen
  \bibfield  {author} {\bibinfo {author} {\bibfnamefont {S.}~\bibnamefont
  {Frixione}}, \bibinfo {author} {\bibfnamefont {P.}~\bibnamefont {Nason}}\
  and\ \bibinfo {author} {\bibfnamefont {C.}~\bibnamefont {Oleari}},\ }\href
  {\doibase 10.1088/1126-6708/2007/11/070} {\bibfield  {journal} {\bibinfo
  {journal} {JHEP}\ }\textbf {\bibinfo {volume} {11}},\ \bibinfo {pages} {070}
  (\bibinfo {year} {2007}{\natexlab{b}})},\ \Eprint
  {http://arxiv.org/abs/0709.2092} {arXiv:0709.2092 [hep-ph]}\BibitemShut
  {NoStop}%
%%CITATION = ARXIV:0709.2092;%%
\bibitem [{\citenamefont {Alioli}\ \emph {et~al.}(2010)\citenamefont {Alioli},
  \citenamefont {Nason}, \citenamefont {Oleari},\ and\ \citenamefont
  {Re}}]{Alioli:2010xd}%
  \BibitemOpen
  \bibfield  {author} {\bibinfo {author} {\bibfnamefont {S.}~\bibnamefont
  {Alioli}}, \bibinfo {author} {\bibfnamefont {P.}~\bibnamefont {Nason}},
  \bibinfo {author} {\bibfnamefont {C.}~\bibnamefont {Oleari}}\ and\ \bibinfo
  {author} {\bibfnamefont {E.}~\bibnamefont {Re}},\ }\href {\doibase
  10.1007/JHEP06(2010)043} {\bibfield  {journal} {\bibinfo  {journal} {JHEP}\
  }\textbf {\bibinfo {volume} {06}},\ \bibinfo {pages} {043} (\bibinfo {year}
  {2010})},\ \Eprint {http://arxiv.org/abs/1002.2581} {arXiv:1002.2581
  [hep-ph]}\BibitemShut {NoStop}%
%%CITATION = ARXIV:1002.2581;%%
\bibitem [{\citenamefont {Je\v{z}o}\ and\ \citenamefont
  {Nason}(2015)}]{Jezo:2015aia}%
  \BibitemOpen
  \bibfield  {author} {\bibinfo {author} {\bibfnamefont {T.}~\bibnamefont
  {Je\v{z}o}}\ and\ \bibinfo {author} {\bibfnamefont {P.}~\bibnamefont
  {Nason}},\ }\href {\doibase 10.1007/JHEP12(2015)065} {\bibfield  {journal}
  {\bibinfo  {journal} {JHEP}\ }\textbf {\bibinfo {volume} {12}},\ \bibinfo
  {pages} {065} (\bibinfo {year} {2015})},\ \Eprint
  {http://arxiv.org/abs/1509.09071} {arXiv:1509.09071 [hep-ph]}\BibitemShut
  {NoStop}%
\bibitem [{\citenamefont {Cascioli}\ \emph {et~al.}(2012)\citenamefont
  {Cascioli}, \citenamefont {Maierh\"ofer},\ and\ \citenamefont
  {Pozzorini}}]{Cascioli:2011va}%
  \BibitemOpen
  \bibfield  {author} {\bibinfo {author} {\bibfnamefont {F.}~\bibnamefont
  {Cascioli}}, \bibinfo {author} {\bibfnamefont {P.}~\bibnamefont
  {Maierh\"ofer}}\ and\ \bibinfo {author} {\bibfnamefont {S.}~\bibnamefont
  {Pozzorini}},\ }\href {\doibase 10.1103/PhysRevLett.108.111601} {\bibfield
  {journal} {\bibinfo  {journal} {Phys. Rev. Lett.}\ }\textbf {\bibinfo
  {volume} {108}},\ \bibinfo {pages} {111601} (\bibinfo {year} {2012})},\
  \Eprint {http://arxiv.org/abs/1111.5206} {arXiv:1111.5206
  [hep-ph]}\BibitemShut {NoStop}%
%%CITATION = ARXIV:1111.5206;%%
\bibitem [{\citenamefont {Buccioni}\ \emph {et~al.}(2018)\citenamefont
  {Buccioni}, \citenamefont {Pozzorini},\ and\ \citenamefont
  {Zoller}}]{Buccioni:2017yxi}%
  \BibitemOpen
  \bibfield  {author} {\bibinfo {author} {\bibfnamefont {F.}~\bibnamefont
  {Buccioni}}, \bibinfo {author} {\bibfnamefont {S.}~\bibnamefont {Pozzorini}}\
  and\ \bibinfo {author} {\bibfnamefont {M.}~\bibnamefont {Zoller}},\ }\href
  {\doibase 10.1140/epjc/s10052-018-5562-1} {\bibfield  {journal} {\bibinfo
  {journal} {Eur. Phys. J.}\ }\textbf {\bibinfo {volume} {C78}},\ \bibinfo
  {pages} {70} (\bibinfo {year} {2018})},\ \Eprint
  {http://arxiv.org/abs/1710.11452} {arXiv:1710.11452 [hep-ph]}\BibitemShut
  {NoStop}%
%%CITATION = ARXIV:1710.11452;%%
\bibitem [{\citenamefont {Buccioni}\ \emph {et~al.}(2019)\citenamefont
  {Buccioni}, \citenamefont {Lang}, \citenamefont {Lindert}, \citenamefont
  {Maierh{\"o}fer}, \citenamefont {Pozzorini}, \citenamefont {Zhang},\ and\
  \citenamefont {Zoller}}]{Buccioni:2019sur}%
  \BibitemOpen
  \bibfield  {author} {\bibinfo {author} {\bibfnamefont {F.}~\bibnamefont
  {Buccioni}}, \bibinfo {author} {\bibfnamefont {J.-N.}\ \bibnamefont {Lang}},
  \bibinfo {author} {\bibfnamefont {J.~M.}\ \bibnamefont {Lindert}}, \bibinfo
  {author} {\bibfnamefont {P.}~\bibnamefont {Maierh{\"o}fer}}, \bibinfo
  {author} {\bibfnamefont {S.}~\bibnamefont {Pozzorini}}, \bibinfo {author}
  {\bibfnamefont {H.}~\bibnamefont {Zhang}}\ and\ \bibinfo {author}
  {\bibfnamefont {M.~F.}\ \bibnamefont {Zoller}},\ }\href {\doibase
  10.1140/epjc/s10052-019-7306-2} {\bibfield  {journal} {\bibinfo  {journal}
  {Eur. Phys. J. C}\ }\textbf {\bibinfo {volume} {79}},\ \bibinfo {pages} {866}
  (\bibinfo {year} {2019})},\ \Eprint {http://arxiv.org/abs/1907.13071}
  {arXiv:1907.13071 [hep-ph]}\BibitemShut {NoStop}%
\bibitem [{\citenamefont {Je\v{z}o}\ \emph {et~al.}(2016)\citenamefont
  {Je\v{z}o}, \citenamefont {Lindert}, \citenamefont {Nason}, \citenamefont
  {Oleari},\ and\ \citenamefont {Pozzorini}}]{Jezo:2016ujg}%
  \BibitemOpen
  \bibfield  {author} {\bibinfo {author} {\bibfnamefont {T.}~\bibnamefont
  {Je\v{z}o}}, \bibinfo {author} {\bibfnamefont {J.~M.}\ \bibnamefont
  {Lindert}}, \bibinfo {author} {\bibfnamefont {P.}~\bibnamefont {Nason}},
  \bibinfo {author} {\bibfnamefont {C.}~\bibnamefont {Oleari}}\ and\ \bibinfo
  {author} {\bibfnamefont {S.}~\bibnamefont {Pozzorini}},\ }\href {\doibase
  10.1140/epjc/s10052-016-4538-2} {\bibfield  {journal} {\bibinfo  {journal}
  {Eur. Phys. J. C}\ }\textbf {\bibinfo {volume} {76}},\ \bibinfo {pages} {691}
  (\bibinfo {year} {2016})},\ \Eprint {http://arxiv.org/abs/1607.04538}
  {arXiv:1607.04538 [hep-ph]}\BibitemShut {NoStop}%
\bibitem [{\citenamefont {B\"arnreuther}\ \emph {et~al.}(2014)\citenamefont
  {B\"arnreuther}, \citenamefont {Czakon},\ and\ \citenamefont
  {Fiedler}}]{Barnreuther:2013qvf}%
  \BibitemOpen
  \bibfield  {author} {\bibinfo {author} {\bibfnamefont {P.}~\bibnamefont
  {B\"arnreuther}}, \bibinfo {author} {\bibfnamefont {M.}~\bibnamefont
  {Czakon}}\ and\ \bibinfo {author} {\bibfnamefont {P.}~\bibnamefont
  {Fiedler}},\ }\href {\doibase 10.1007/JHEP02(2014)078} {\bibfield  {journal}
  {\bibinfo  {journal} {JHEP}\ }\textbf {\bibinfo {volume} {02}},\ \bibinfo
  {pages} {078} (\bibinfo {year} {2014})},\ \Eprint
  {http://arxiv.org/abs/1312.6279} {arXiv:1312.6279 [hep-ph]}\BibitemShut
  {NoStop}%
\bibitem [{\citenamefont {Ball}\ \emph {et~al.}(2017)\citenamefont {Ball} \emph
  {et~al.}}]{Ball:2017nwa}%
  \BibitemOpen
  \bibfield  {author} {\bibinfo {author} {\bibfnamefont {R.~D.}\ \bibnamefont
  {Ball}} \emph {et~al.} (\bibinfo {collaboration} {NNPDF}),\ }\href {\doibase
  10.1140/epjc/s10052-017-5199-5} {\bibfield  {journal} {\bibinfo  {journal}
  {Eur. Phys. J.}\ }\textbf {\bibinfo {volume} {C77}},\ \bibinfo {pages} {663}
  (\bibinfo {year} {2017})},\ \Eprint {http://arxiv.org/abs/1706.00428}
  {arXiv:1706.00428 [hep-ph]}\BibitemShut {NoStop}%
%%CITATION = ARXIV:1706.00428;%%
\bibitem [{\citenamefont {Buckley}\ \emph {et~al.}(2015)\citenamefont
  {Buckley}, \citenamefont {Ferrando}, \citenamefont {Lloyd}, \citenamefont
  {Nordström}, \citenamefont {Page}, \citenamefont {Rüfenacht}, \citenamefont
  {Schönherr},\ and\ \citenamefont {Watt}}]{Buckley:2014ana}%
  \BibitemOpen
  \bibfield  {author} {\bibinfo {author} {\bibfnamefont {A.}~\bibnamefont
  {Buckley}}, \bibinfo {author} {\bibfnamefont {J.}~\bibnamefont {Ferrando}},
  \bibinfo {author} {\bibfnamefont {S.}~\bibnamefont {Lloyd}}, \bibinfo
  {author} {\bibfnamefont {K.}~\bibnamefont {Nordström}}, \bibinfo {author}
  {\bibfnamefont {B.}~\bibnamefont {Page}}, \bibinfo {author} {\bibfnamefont
  {M.}~\bibnamefont {Rüfenacht}}, \bibinfo {author} {\bibfnamefont
  {M.}~\bibnamefont {Schönherr}}\ and\ \bibinfo {author} {\bibfnamefont
  {G.}~\bibnamefont {Watt}},\ }\href {\doibase 10.1140/epjc/s10052-015-3318-8}
  {\bibfield  {journal} {\bibinfo  {journal} {Eur. Phys. J.}\ }\textbf
  {\bibinfo {volume} {C75}},\ \bibinfo {pages} {132} (\bibinfo {year}
  {2015})},\ \Eprint {http://arxiv.org/abs/1412.7420} {arXiv:1412.7420
  [hep-ph]}\BibitemShut {NoStop}%
%%CITATION = ARXIV:1412.7420;%%
\bibitem [{\citenamefont {Salam}\ and\ \citenamefont
  {Rojo}(2009)}]{Salam:2008qg}%
  \BibitemOpen
  \bibfield  {author} {\bibinfo {author} {\bibfnamefont {G.~P.}\ \bibnamefont
  {Salam}}\ and\ \bibinfo {author} {\bibfnamefont {J.}~\bibnamefont {Rojo}},\
  }\href {\doibase 10.1016/j.cpc.2008.08.010} {\bibfield  {journal} {\bibinfo
  {journal} {Comput. Phys. Commun.}\ }\textbf {\bibinfo {volume} {180}},\
  \bibinfo {pages} {120} (\bibinfo {year} {2009})},\ \Eprint
  {http://arxiv.org/abs/0804.3755} {arXiv:0804.3755 [hep-ph]}\BibitemShut
  {NoStop}%
%%CITATION = ARXIV:0804.3755;%%
\bibitem [{\citenamefont {Sjöstrand}\ \emph {et~al.}(2015)\citenamefont
  {Sjöstrand}, \citenamefont {Ask}, \citenamefont {Christiansen},
  \citenamefont {Corke}, \citenamefont {Desai}, \citenamefont {Ilten},
  \citenamefont {Mrenna}, \citenamefont {Prestel}, \citenamefont {Rasmussen},\
  and\ \citenamefont {Skands}}]{Sjostrand:2014zea}%
  \BibitemOpen
  \bibfield  {author} {\bibinfo {author} {\bibfnamefont {T.}~\bibnamefont
  {Sjöstrand}}, \bibinfo {author} {\bibfnamefont {S.}~\bibnamefont {Ask}},
  \bibinfo {author} {\bibfnamefont {J.~R.}\ \bibnamefont {Christiansen}},
  \bibinfo {author} {\bibfnamefont {R.}~\bibnamefont {Corke}}, \bibinfo
  {author} {\bibfnamefont {N.}~\bibnamefont {Desai}}, \bibinfo {author}
  {\bibfnamefont {P.}~\bibnamefont {Ilten}}, \bibinfo {author} {\bibfnamefont
  {S.}~\bibnamefont {Mrenna}}, \bibinfo {author} {\bibfnamefont
  {S.}~\bibnamefont {Prestel}}, \bibinfo {author} {\bibfnamefont {C.~O.}\
  \bibnamefont {Rasmussen}}\ and\ \bibinfo {author} {\bibfnamefont {P.~Z.}\
  \bibnamefont {Skands}},\ }\href {\doibase 10.1016/j.cpc.2015.01.024}
  {\bibfield  {journal} {\bibinfo  {journal} {Comput. Phys. Commun.}\ }\textbf
  {\bibinfo {volume} {191}},\ \bibinfo {pages} {159} (\bibinfo {year}
  {2015})},\ \Eprint {http://arxiv.org/abs/1410.3012} {arXiv:1410.3012
  [hep-ph]}\BibitemShut {NoStop}%
%%CITATION = ARXIV:1410.3012;%%
\bibitem [{\citenamefont {Skands}\ \emph {et~al.}(2014)\citenamefont {Skands},
  \citenamefont {Carrazza},\ and\ \citenamefont {Rojo}}]{Skands:2014pea}%
  \BibitemOpen
  \bibfield  {author} {\bibinfo {author} {\bibfnamefont {P.}~\bibnamefont
  {Skands}}, \bibinfo {author} {\bibfnamefont {S.}~\bibnamefont {Carrazza}}\
  and\ \bibinfo {author} {\bibfnamefont {J.}~\bibnamefont {Rojo}},\ }\href
  {\doibase 10.1140/epjc/s10052-014-3024-y} {\bibfield  {journal} {\bibinfo
  {journal} {Eur. Phys. J. C}\ }\textbf {\bibinfo {volume} {74}},\ \bibinfo
  {pages} {3024} (\bibinfo {year} {2014})},\ \Eprint
  {http://arxiv.org/abs/1404.5630} {arXiv:1404.5630 [hep-ph]}\BibitemShut
  {NoStop}%
\bibitem [{\citenamefont {Banfi}\ \emph {et~al.}(2006)\citenamefont {Banfi},
  \citenamefont {Salam},\ and\ \citenamefont {Zanderighi}}]{Banfi:2006hf}%
  \BibitemOpen
  \bibfield  {author} {\bibinfo {author} {\bibfnamefont {A.}~\bibnamefont
  {Banfi}}, \bibinfo {author} {\bibfnamefont {G.~P.}\ \bibnamefont {Salam}}\
  and\ \bibinfo {author} {\bibfnamefont {G.}~\bibnamefont {Zanderighi}},\
  }\href {\doibase 10.1140/epjc/s2006-02552-4} {\bibfield  {journal} {\bibinfo
  {journal} {Eur. Phys. J. C}\ }\textbf {\bibinfo {volume} {47}},\ \bibinfo
  {pages} {113} (\bibinfo {year} {2006})},\ \Eprint
  {http://arxiv.org/abs/hep-ph/0601139} {arXiv:hep-ph/0601139}\BibitemShut
  {NoStop}%
\bibitem [{\citenamefont {Buckley}\ and\ \citenamefont
  {Pollard}(2016)}]{Buckley:2015gua}%
  \BibitemOpen
  \bibfield  {author} {\bibinfo {author} {\bibfnamefont {A.}~\bibnamefont
  {Buckley}}\ and\ \bibinfo {author} {\bibfnamefont {C.}~\bibnamefont
  {Pollard}},\ }\href {\doibase 10.1140/epjc/s10052-016-3925-z} {\bibfield
  {journal} {\bibinfo  {journal} {Eur. Phys. J. C}\ }\textbf {\bibinfo {volume}
  {76}},\ \bibinfo {pages} {71} (\bibinfo {year} {2016})},\ \Eprint
  {http://arxiv.org/abs/1507.00508} {arXiv:1507.00508 [hep-ph]}\BibitemShut
  {NoStop}%
\bibitem [{\citenamefont {Ilten}\ \emph {et~al.}(2017)\citenamefont {Ilten},
  \citenamefont {Rodd}, \citenamefont {Thaler},\ and\ \citenamefont
  {Williams}}]{Ilten:2017rbd}%
  \BibitemOpen
  \bibfield  {author} {\bibinfo {author} {\bibfnamefont {P.}~\bibnamefont
  {Ilten}}, \bibinfo {author} {\bibfnamefont {N.~L.}\ \bibnamefont {Rodd}},
  \bibinfo {author} {\bibfnamefont {J.}~\bibnamefont {Thaler}}\ and\ \bibinfo
  {author} {\bibfnamefont {M.}~\bibnamefont {Williams}},\ }\href {\doibase
  10.1103/PhysRevD.96.054019} {\bibfield  {journal} {\bibinfo  {journal} {Phys.
  Rev. D}\ }\textbf {\bibinfo {volume} {96}},\ \bibinfo {pages} {054019}
  (\bibinfo {year} {2017})},\ \Eprint {http://arxiv.org/abs/1702.02947}
  {arXiv:1702.02947 [hep-ph]}\BibitemShut {NoStop}%
\bibitem [{\citenamefont {Caletti}\ \emph {et~al.}(2021)\citenamefont
  {Caletti}, \citenamefont {Fedkevych}, \citenamefont {Marzani}, \citenamefont
  {Reichelt}, \citenamefont {Schumann}, \citenamefont {Soyez},\ and\
  \citenamefont {Theeuwes}}]{Caletti:2021oor}%
  \BibitemOpen
  \bibfield  {author} {\bibinfo {author} {\bibfnamefont {S.}~\bibnamefont
  {Caletti}}, \bibinfo {author} {\bibfnamefont {O.}~\bibnamefont {Fedkevych}},
  \bibinfo {author} {\bibfnamefont {S.}~\bibnamefont {Marzani}}, \bibinfo
  {author} {\bibfnamefont {D.}~\bibnamefont {Reichelt}}, \bibinfo {author}
  {\bibfnamefont {S.}~\bibnamefont {Schumann}}, \bibinfo {author}
  {\bibfnamefont {G.}~\bibnamefont {Soyez}}\ and\ \bibinfo {author}
  {\bibfnamefont {V.}~\bibnamefont {Theeuwes}},\ }\href {\doibase
  10.1007/JHEP07(2021)076} {\bibfield  {journal} {\bibinfo  {journal} {JHEP}\
  }\textbf {\bibinfo {volume} {07}},\ \bibinfo {pages} {076} (\bibinfo {year}
  {2021})},\ \Eprint {http://arxiv.org/abs/2104.06920} {arXiv:2104.06920
  [hep-ph]}\BibitemShut {NoStop}%
\bibitem [{\citenamefont {Fedkevych}\ \emph {et~al.}(2022)\citenamefont
  {Fedkevych}, \citenamefont {Khosa}, \citenamefont {Marzani},\ and\
  \citenamefont {Sforza}}]{Fedkevych:2022mid}%
  \BibitemOpen
  \bibfield  {author} {\bibinfo {author} {\bibfnamefont {O.}~\bibnamefont
  {Fedkevych}}, \bibinfo {author} {\bibfnamefont {C.~K.}\ \bibnamefont
  {Khosa}}, \bibinfo {author} {\bibfnamefont {S.}~\bibnamefont {Marzani}}\ and\
  \bibinfo {author} {\bibfnamefont {F.}~\bibnamefont {Sforza}},\ }\href@noop {}
  {\  (\bibinfo {year} {2022})},\ \Eprint {http://arxiv.org/abs/2202.05082}
  {arXiv:2202.05082 [hep-ph]}\BibitemShut {NoStop}%
\bibitem [{\citenamefont {Caletti}\ \emph
  {et~al.}(2022{\natexlab{a}})\citenamefont {Caletti}, \citenamefont
  {Larkoski}, \citenamefont {Marzani},\ and\ \citenamefont
  {Reichelt}}]{Caletti:2022hnc}%
  \BibitemOpen
  \bibfield  {author} {\bibinfo {author} {\bibfnamefont {S.}~\bibnamefont
  {Caletti}}, \bibinfo {author} {\bibfnamefont {A.~J.}\ \bibnamefont
  {Larkoski}}, \bibinfo {author} {\bibfnamefont {S.}~\bibnamefont {Marzani}}\
  and\ \bibinfo {author} {\bibfnamefont {D.}~\bibnamefont {Reichelt}},\ }\href
  {\doibase 10.1140/epjc/s10052-022-10568-7} {\bibfield  {journal} {\bibinfo
  {journal} {Eur. Phys. J. C}\ }\textbf {\bibinfo {volume} {82}},\ \bibinfo
  {pages} {632} (\bibinfo {year} {2022}{\natexlab{a}})},\ \Eprint
  {http://arxiv.org/abs/2205.01109} {arXiv:2205.01109 [hep-ph]}\BibitemShut
  {NoStop}%
\bibitem [{\citenamefont {Caletti}\ \emph
  {et~al.}(2022{\natexlab{b}})\citenamefont {Caletti}, \citenamefont
  {Larkoski}, \citenamefont {Marzani},\ and\ \citenamefont
  {Reichelt}}]{Caletti:2022glq}%
  \BibitemOpen
  \bibfield  {author} {\bibinfo {author} {\bibfnamefont {S.}~\bibnamefont
  {Caletti}}, \bibinfo {author} {\bibfnamefont {A.~J.}\ \bibnamefont
  {Larkoski}}, \bibinfo {author} {\bibfnamefont {S.}~\bibnamefont {Marzani}}\
  and\ \bibinfo {author} {\bibfnamefont {D.}~\bibnamefont {Reichelt}},\ }\href
  {\doibase 10.1007/JHEP10(2022)158} {\bibfield  {journal} {\bibinfo  {journal}
  {JHEP}\ }\textbf {\bibinfo {volume} {10}},\ \bibinfo {pages} {158} (\bibinfo
  {year} {2022}{\natexlab{b}})},\ \Eprint {http://arxiv.org/abs/2205.01117}
  {arXiv:2205.01117 [hep-ph]}\BibitemShut {NoStop}%
\bibitem [{\citenamefont {Czakon}\ \emph {et~al.}(2022)\citenamefont {Czakon},
  \citenamefont {Mitov},\ and\ \citenamefont {Poncelet}}]{Czakon:2022wam}%
  \BibitemOpen
  \bibfield  {author} {\bibinfo {author} {\bibfnamefont {M.}~\bibnamefont
  {Czakon}}, \bibinfo {author} {\bibfnamefont {A.}~\bibnamefont {Mitov}}\ and\
  \bibinfo {author} {\bibfnamefont {R.}~\bibnamefont {Poncelet}},\ }\href@noop
  {} {\  (\bibinfo {year} {2022})},\ \Eprint {http://arxiv.org/abs/2205.11879}
  {arXiv:2205.11879 [hep-ph]}\BibitemShut {NoStop}%
\bibitem [{\citenamefont {Gauld}\ \emph {et~al.}(2022)\citenamefont {Gauld},
  \citenamefont {Huss},\ and\ \citenamefont {Stagnitto}}]{Gauld:2022lem}%
  \BibitemOpen
  \bibfield  {author} {\bibinfo {author} {\bibfnamefont {R.}~\bibnamefont
  {Gauld}}, \bibinfo {author} {\bibfnamefont {A.}~\bibnamefont {Huss}}\ and\
  \bibinfo {author} {\bibfnamefont {G.}~\bibnamefont {Stagnitto}},\ }\href@noop
  {} {\  (\bibinfo {year} {2022})},\ \Eprint {http://arxiv.org/abs/2208.11138}
  {arXiv:2208.11138 [hep-ph]}\BibitemShut {NoStop}%
\end{thebibliography}%

\clearpage

\end{document}